\documentclass[nofootinbib,aps,prd,reprint,superscriptaddress,showkeys]{revtex4-2}

\usepackage[utf8]{inputenc} 
\usepackage{graphicx,overpic,mathtools}
\usepackage{amsthm,amsmath,amssymb,hyperref}
\usepackage{braket,bm,bbm,setspace}
\usepackage[normalem]{ulem} 
\usepackage{physics}
\usepackage{float}
\usepackage[makeroom]{cancel}
\usepackage[english]{babel}
\usepackage{xcolor}
\usepackage{tensor}
\graphicspath{ {images/} }
\addto\captionsspanish{}
\hypersetup{
	colorlinks=true,
	pdfborder={0 0 0},
	citecolor=purple,
	linkcolor=blue,
	filecolor=blue,
	urlcolor=black,
}
\usepackage{subfigure}

\makeatletter
\def\l@subsection#1#2{}
\def\l@subsubsection#1#2{}
\makeatother

\begin{document}
\title{Discrete-time Semiclassical Szegedy Quantum Walks}
\author{Sergio A. Ortega}
%\author{Sergio A. Ortega \orcidlink{0000-0002-8237-7711}}
\email{sergioan@ucm.es}
\affiliation{Departamento de Física Teórica, Universidad Complutense de Madrid, 28040 Madrid, Spain}
\author{Miguel A. Martin-Delgado}
%\author{Miguel A. Martin-Delgado \orcidlink{0000-0003-2746-5062}}
\email{mardel@ucm.es}
\affiliation{Departamento de Física Teórica, Universidad Complutense de Madrid, 28040 Madrid, Spain}
\affiliation{CCS-Center for Computational Simulation, Universidad Politécnica de Madrid, 28660 Boadilla del Monte, Madrid, Spain.}

\begin{abstract}
	{Quantum walks are promising tools based on classical random walks, with plenty of applications such as many variants of optimization. Here we introduce the semiclassical walks in discrete time, which are algorithms that combines classical and quantum dynamics. Specifically, a semiclassical walk can be understood as a classical walk where the transition matrix encodes the quantum evolution. We have applied this algorithm to Szegedy's quantum walk, which can be applied to any arbitrary weighted graph. We first have solved the problem analytically on regular 1D cycles to show the performance of the semiclassical walks. Next, we have simulated our algorithm in a general inhomogeneous symmetric graph, finding that the inhomogeneity drives a symmetry breaking on the graph. Moreover, we show that this phenomenon is useful for the problem of ranking nodes in symmetric graphs, where the classical PageRank fails. We have demonstrated experimentally that the semiclassical walks can be applied on real quantum computers using the platform IBM Quantum.}
\end{abstract}


\maketitle

\section{Introduction}\label{Introduction}

Quantum walks are algorithms born from the quantization of classical random walks. They were first proposed in \cite{QRW} in the discrete time version, and later in \cite{Trees} using a continuous time. However, precursor ideas can be attributed to Feynman \cite{Portugal}.  These walks have given rise to a wide variety of algorithms for problems such us triangle finding \cite{Triangles}, element distinctness \cite{ED} and quantum search \cite{QRW_Search}. Moreover, these algorithms are very interesting because they can simulate a lot of physical systems \cite{Portugal}, and can be used for universal quantum computation \cite{Universal}.

In this paper we want to introduce the concept of semiclassical walk, which is a type algorithm that combines classical and quantum dynamics. The idea of mixing both dynamics is not new. In \cite{QSW} it was introduced the quantum stochastic walk, a parameterized walk driven by non-unitary evolution that interpolates between the quantum and the classical dynamics. Another idea is measuring the position of the walker at regular intervals of time, and let the system evolve with the unitary quantum evolution between the measurements. This algorithm was introduced in \cite{MIQW} using a quantum walk in continuous time, and it was called measurement-induced quantum walk. We aim to generalize this algorithm with measurements to the context where the quantum evolution occurs in discrete time. As we will show in this work, in this scenario it is necessary to reset part of the system after the measurements, so that the system must be controlled by a classical computer beyond the usual quantum gates.

An important quantum walk is the one introduced by Szegedy in \cite{Szegedy} as a generalization of the Grover algorithm \cite{Grover}. In contrast to other approaches, this quantum walk can quantize a general Markov chain, so it can be used over arbitrary weighted graphs. It has been shown to be useful for problems of optimization \cite{Lemieux,Qfold,QMS,GWQMA}, classification \cite{Paparo1,Paparo2,APR}, quantum search \cite{Portugal,Searchrank,S_queries} and machine learning \cite{Paparo3}. Moreover, there has been research in implementing this algorithm in quantum circuits \cite{Q_circuits}. Due to its potential applications, we have chosen this quantum walk as an interesting example for implementing the semiclassical walks. We expect this could give rise to novel algorithms in the future.

This paper is structured as follows. In Section \ref{Formulation} we review the formulation of classical and quantum walks, to later introduce the semiclassical walks in discrete time. In Section \ref{Szegedy} we focus on the semiclassical walks built from Szegedy's quantum walk. In Section \ref{Cycles} we solve analytically the problem for 1D cycles and show results in some examples. In section \ref{Symmetry} we simulate the semiclassical walks in a generic weighted graph, showing how this approach can break the symmetry of the graph. In Section \ref{MIQW} we compare our results with the previous approach in continuous time. In section \ref{Experimental} we show experimental results of semiclassical walks in a real quantum computer. Finally, we summarize and conclude in Section \ref{Conclusions}.

\section{Semiclassical Walk Formulation}\label{Formulation}

In this section we are going to review the formulation of classical and quantum walks in discrete time, to later join them together to build up the semiclassical walk.

\subsection{Classical Walk}

\begin{figure*}
	\centering
	\subfigure[]{\includegraphics[width=0.3525\linewidth]{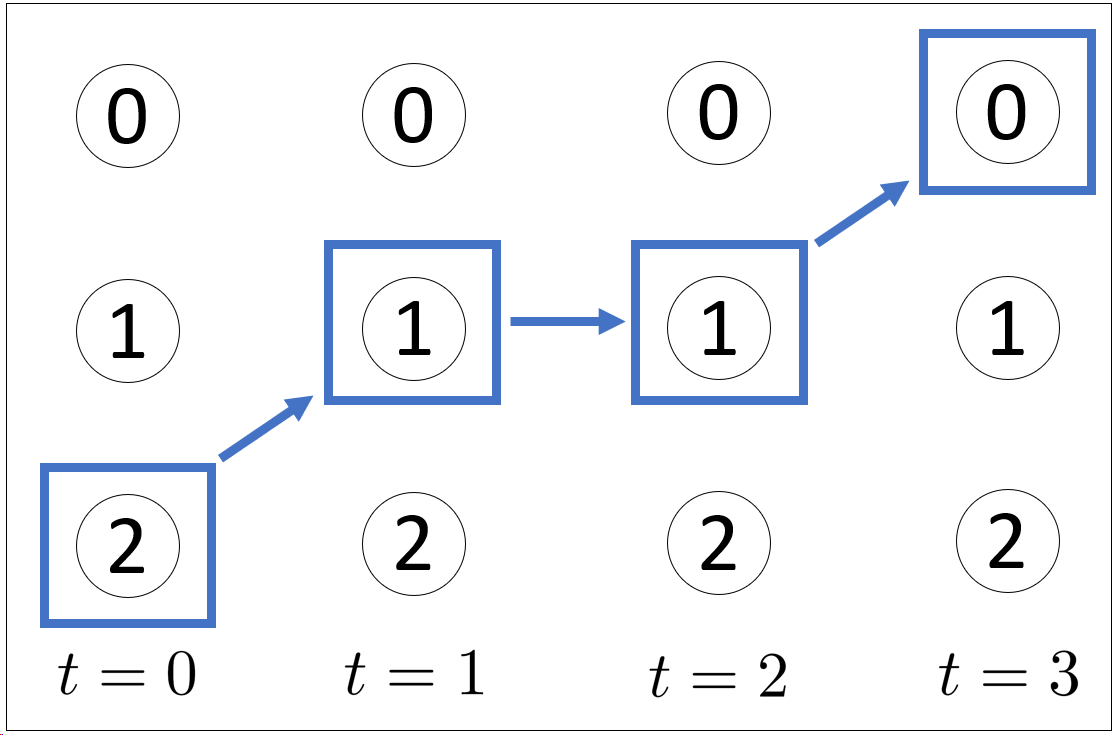}\label{F:classical}}
	\subfigure[]{\includegraphics[width=0.4255\linewidth]{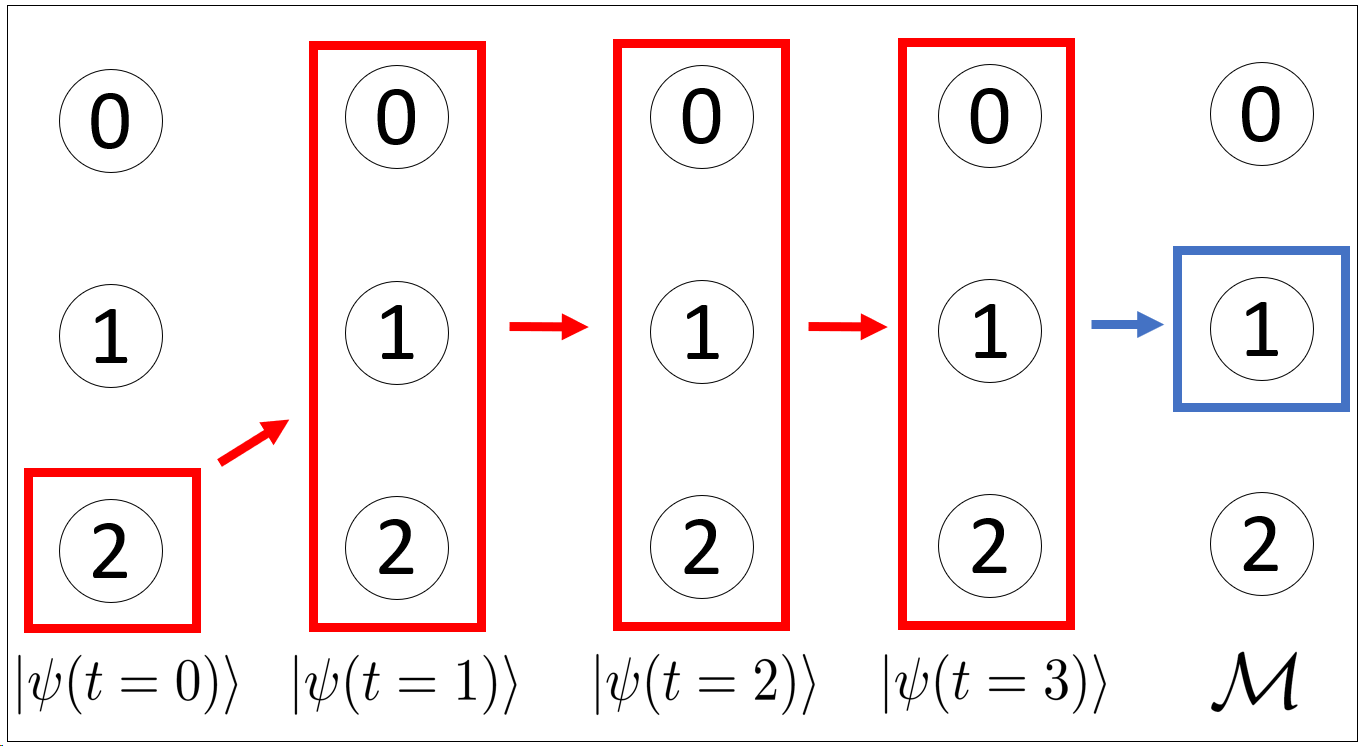}\label{F:quantum}}
	\subfigure[]{\includegraphics[width=0.70\linewidth]{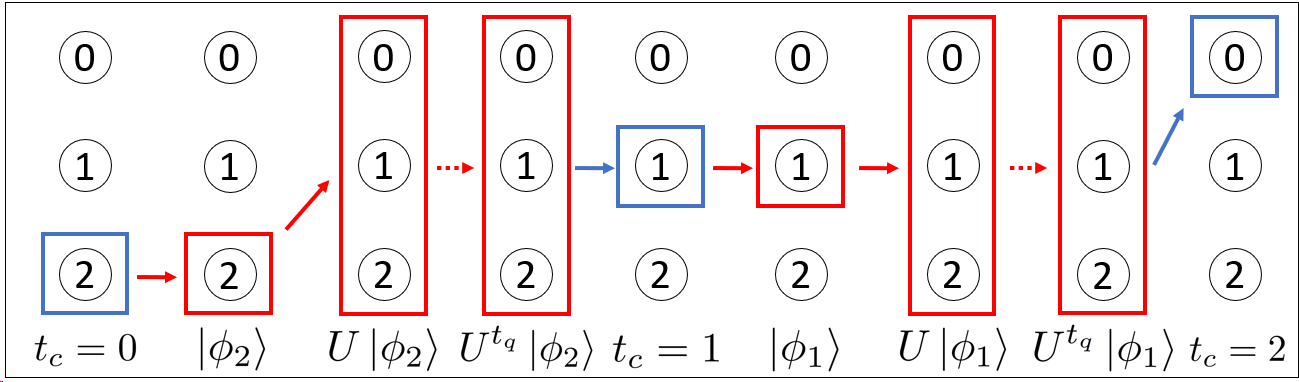}\label{F:semiclassical_1}}
	\subfigure[]{\includegraphics[width=0.2336\linewidth]{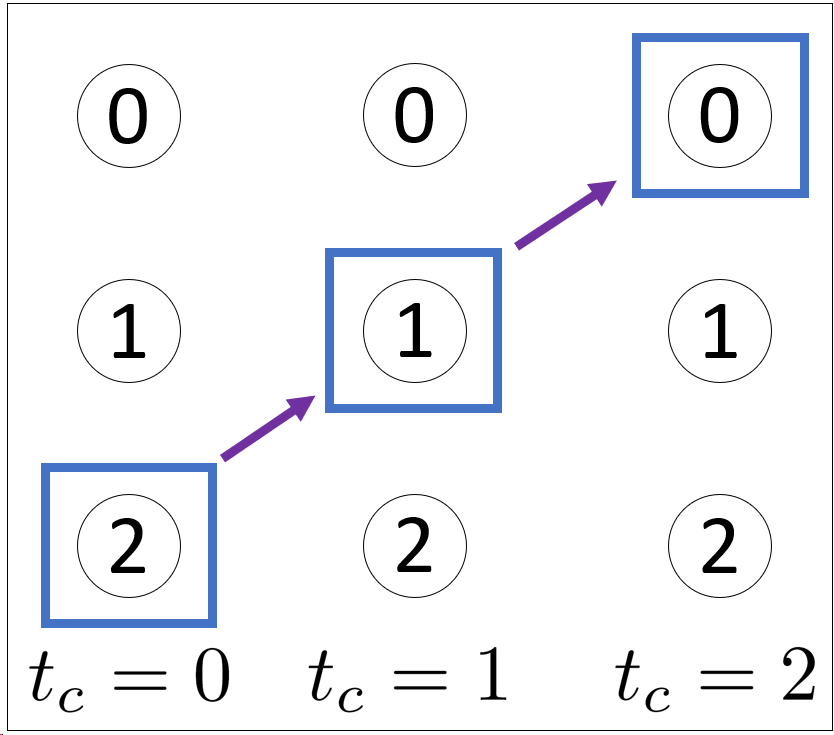}\label{F:semiclassical_2}}
	\caption{Examples of trajectories followed by a particle in the different types of walks in a graph with 3 nodes. Blue color represents classical information whereas red color represents quantum information. a) Classical walk. The particle is in a particular node at each time step, and it jumps to other node with a certain probability. b) Quantum walk. The walker can be in a superposition of the nodes at each time step, and it is represented by a quantum state. In the end of the quantum walk the position is measured so that the walker collapses to a particular state with a certain probability. c) Semiclassical walk. At each classical step the walker is in a particular classical position. For each of these steps, the proxy quantum state is prepared and it is performed the quantum evolution $t_q$ times. After that, the position is measured obtaining a new classical position. d) Representation of the semiclassical trajectory in c) as a classical walk. Purples arrows indicate that the evolution is affected by both classical and quantum dynamics.}
	\label{...}
\end{figure*}

The classical random walk occurs in the nodes of a graph. From a stochastic point of view, at each time step the walker is in only one node of the graph, and can jump to any other node, including the same node, with some probabilities. In Figure \ref{F:classical} it is shown an example of classical walk in a graph with three nodes. At the initial time $t=0$ the walker is at node $2$. At the first time step, the walker decides stochastically to jump to node $1$. At the second time step it remains at node $1$, and at the third time step it jumps to node $0$. Since this process is stochastic, each time we repeat the random walk the trajectory followed by the walker is different. Averaging over the different trajectories we could obtain a probability distribution for the walker being at each node at each time step.

If we knew beforehand the probability that the walker jumps from one node to any other, then we could simulate deterministically the probability distribution of the walker being at each node. Let us define the classical transition matrix of a graph $G$, whose elements $G_{ji}$ are the probabilities of the walker jumping from node $i$ to node $j$. This matrix is column-stochastic by definition, so that all the columns add up to one. Thus
\begin{equation}\label{column_stochastic}
\sum_{j=0}^{N-1} G_{ji} = 1,
\end{equation}
where $N$ is the number of nodes in the graph. In this paper we count the nodes of the network, and therefore the matrix indexes, from $0$ to $N-1$. Let $p(t)$ be a column vector whose elements are the probabilities of the walker being at each node at time $t$. Then, given an initial probability distribution $p(0)$, the classical walk can be simulated deterministically as
\begin{equation}\label{classical_evolution}
p(t) = G^t p(0).
\end{equation}

Both points of view for the classical walk have applications depending on whether we want to obtain a specific position of the walker or the probability distribution. For example, the stochastic simulation is used for optimization algorithms like simulated annealing with Metropolis-Hastings, where it is wanted to obtain a single node as an optimal solution \cite{Metropolis,Hastings,SimAnn}. The search space can be so big that it is unfeasible to calculate the entire transition matrix. Thus, only the probabilities needed at each time step are calculated, reducing the computational cost of the algorithm. Other example is the PageRank algorithm, where the objective is the limiting probability distribution of the walker for classifying the nodes of the graph \cite{Brin1,Brin2,Brin3,Google_book}.

\subsection{Quantum Walk}

A quantum walk is a quantization of a classical walk. In this case, the walker can be in a superposition of the nodes of the graph, so that the position at each time step is represented by a quantum state $\left|\psi(t)\right>$. The evolution of the system is given by a unitary matrix $U$, so that
\begin{equation}
\left|\psi(t)\right> = U^t \left|\psi(0)\right>.
\end{equation}
Let us define the computational basis as the one formed by the vectors $\left|i\right>$, with $i = 0,1,...,N-1$ representing the nodes of the graph. Then, the probability that the quantum walker can be measured at each node at time step $t$ is:
\begin{equation}
p_i(t) = \left|\left|\left<i|\psi(t)\right>\right|\right|^2.
\end{equation}
An example of a quantum walk in a graph with three nodes is shown in Figure \ref{F:quantum}. The walker start in a quantum state that represents node $2$, it performs three time steps of the walk, being in a superposition of all nodes, and finally we measure the position obtaining node $1$.

If we knew the unitary matrix $U$, then we could simulate deterministically the quantum walk in a classical computer, obtaining the probability distribution at each time step. This would be equivalent to the deterministic point of view of the classical walk. However, in some cases it is very costly to simulate the quantum walk, so that it is thought to be performed in quantum hardware instead. In this real scenario, after measuring the quantum state collapses to a state of the computational basis with a certain probability, being analog to the stochastic simulation of the classical walk. This is useful for example in the case of the quantum Metropolis algorithm \cite{Lemieux,Qfold,QMS,GWQMA}. If we wanted the probability distribution we would have to repeat the walk several times and average the results. Moreover, since the state collapses after the measurement, we cannot measure at intermediate steps of the walk and resume it. If we wanted the probability distribution at each time step we would have to perform a different quantum walk for each final time. This is for example what is done in the quantum version of the PageRank algorithm \cite{Paparo1,Paparo2,APR}.

\subsection{Semiclassical Walk}

As we have mentioned above, if we measure the position of the quantum walker at a certain time, the state collapses so that the coherence of the quantum state is broken. We wonder what happens if we resume the quantum evolution after the measurement. As we will see, this new walk can be expressed as a classical walk with a different transition matrix. Let us define the following quantities:

-$t_q$: Quantum time \cite{QT}. It is the number of times we apply the unitary evolution $U$ between measurements.

-$t_c$: Classical time. It is the number of times we apply the quantum evolution $U^{t_q}$ and measure the position of the walker.

Let us define the states $\left|\phi_i\right>$ as proxies for the positions $i$ of the graph. Thus, if the walker is at node $i$ at each classical time $t_c$, then we prepare the state $\left|\phi_i\right>$ to perform the quantum evolution $U$ $t_q$ times. An example is shown in Figure \ref{F:semiclassical_1}. There, the walker start at node $2$. We prepare the quantum proxy state $\left|\phi_2\right>$ and perform the quantum evolution $t_q$ times. After measuring the position we obtain that it is at node $1$. This corresponds to the first classical step $t_c = 1$. For the second classical step $t_c=2$, we prepare the quantum proxy $\left|\phi_1\right>$, perform the quantum evolution $t_q$ times, and measure, obtaining that the walker is at node $0$. If we treat the quantum evolution as a black box and we only deal with the positions after each measurement, then we can treat the walk as a classical walk as shown in Figure \ref{F:semiclassical_2}. Thus, we only see that the walker starts at node $2$, jumps to node $1$, and after that it jumps to node $0$.

As in the classical walk, a particular trajectory of the walker is obtained with a certain probability each time we run the algorithm. If we knew the probability of measuring each node after the quantum evolution starting with the proxy state $\left|\phi_i\right>$, then we could define a transition matrix and simulate the walk in a similar way as equation \eqref{classical_evolution}. Thus, let us define the semiclassical transition matrix for a semiclassical walk as $G^{\left(t_q\right)}$, whose elements are
\begin{equation}
G_{ji}^{\left(t_q\right)} := \left|\left|\left<j|U^{t_q}|\phi_i\right>\right|\right|^2.
\end{equation}
Note that there is a different semiclassical matrix for each value of $t_q$, that is, there is a different semiclassical walk for each number of times we perform the quantum unitary evolution $U$ between measurements. Thus, we actually have a family of semiclassical walks. The quantum time $t_q$ is actually a parameter that defines a particular semiclassical walk in the family, whereas the classical time $t_c$ is the actual evolution time, since we would only deal with the particle position at each classical step, and not at intermediary steps of the quantum evolution.

Finally, we have to define how to construct the proxy states $\left|\phi_i\right>$. A priori they could be exactly the same as the computational basis states $\left|i\right>$, so that when we measure the position of the system, it collapses to the proxy state and it is ready for the next quantum evolution. That would be true if the Hilbert space where the quantum evolution takes place were the span of the states $\left|i\right>$. For a quantum walk in continuous time this is true, and this is how the measurement-induced quantum walk studied in \cite{MIQW} is performed. However, for discrete time quantum walks the space have to be extended with an auxiliary register, usually called coin register, that tells the possible jumps of the quantum walker at each position. This prevents us from constructing a semiclassical walk only measuring the position state. Suppose we start at node $3$, with the coin register in the state $\left|c\right>$. The total wave function would be
\begin{equation}
\left|\psi\right> = \left|3\right> \otimes \left|c\right>.
\end{equation}
Now, suppose we evolve the system with the unitary $U$ $t_q$ times and measure it, obtaining again node $3$. Then the system would collapse to the following quantum state:
\begin{equation}
\left|\psi'\right> = \left|3\right> \otimes \left|c'\right>.
\end{equation}
In general, the coin state after the projection is not the same as in the initial state, even if we measured the same initial node. Since the probabilities of measuring each node after the quantum evolution depends not only on the initial position but also on the initial coin state, then the next quantum evolution would not provide the same probabilities. Thus, the process would not be Markovian, since the state at $t_c = 2$, would depend on the state of the coin at $t_c = 1$, which in turn depends on the state at $t_c=0$.

To overcome this issue we propose a reset scheme, so that with the information of the position after the measurement, we delete the information of the coin register and reprepare it in the initial coin state. In other words, we chose the proxy states as
\begin{equation}\label{proxies}
\left|\phi_i\right> := \left|i\right> \otimes \left|c_i\right>,
\end{equation}
where $c_i$ is a coin state that can be different for each node in the network. Thus, after the measurement, we prepare the proxy $\left|\phi_i\right>$ using only the information of the position register, without taking into account the coin register. In general, the coin states in the proxies can be chosen arbitrarily, giving rise to different semiclassical walk families. However, for Szegedy's quantum walk we will see in the next section that there is a natural definition for them.

\section{Szegedy Semiclassical Walk}\label{Szegedy}

Szegedy's quantum walk was proposed in \cite{Szegedy} as a coinless quantization of Markov chains. In contrast to previous coined quantum walks, which were only suitable for unweighted graphs, Szegedy's quantum walk is able to quantize any classical transition matrix $G$.

\begin{figure*}
	\centering
	\subfigure[]{\includegraphics[width=1\linewidth]{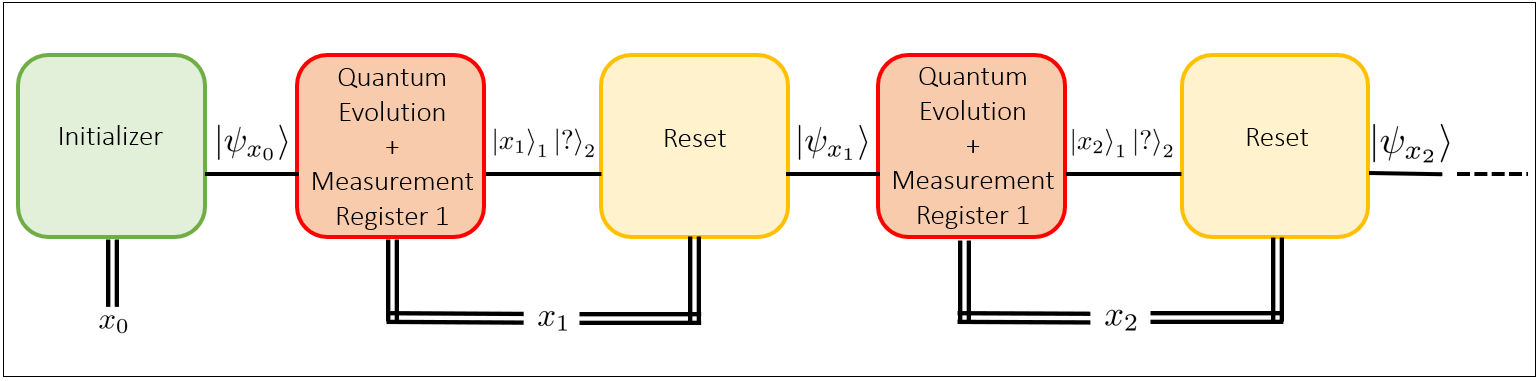}\label{F:Szegedy_sc_1}}
	\subfigure[]{\includegraphics[width=1\linewidth]{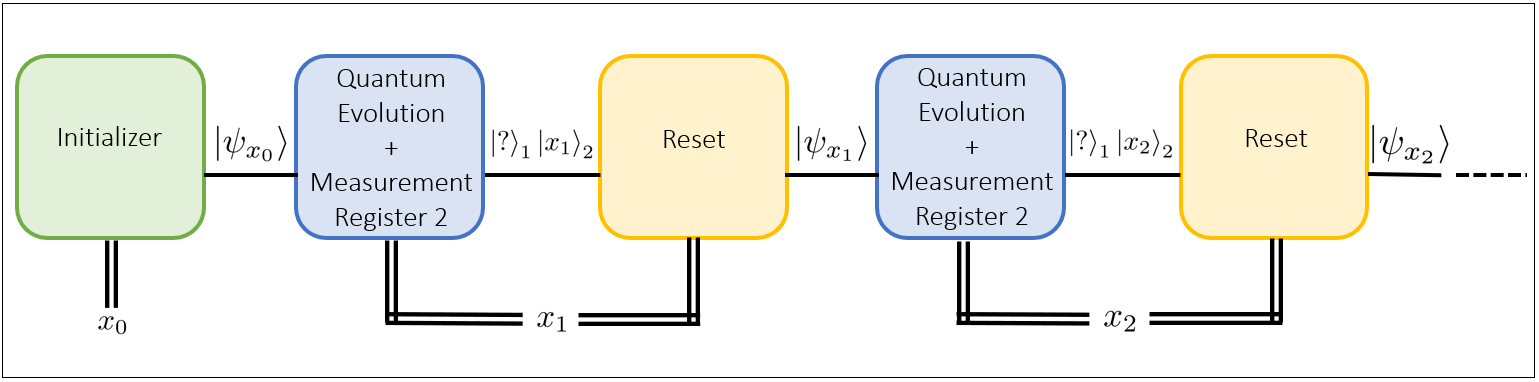}\label{F:Szegedy_sc_2}}
	\caption{a) Scheme of the semiclassical Szegedy's walk of class I. The position of the walker at each classical time step is represented by $x_{t_c}$. The classical information is used to prepare the corresponding proxy state, the quantum evolution is performed, and finally a new classical position is measured from the first register. b) Scheme of the semiclassical Szegedy's walk of class II. In this case, the second register is measured, and that information is used to prepare the new proxy state.}
	\label{F:S_circuit}
\end{figure*}

In Szegedy's quantum walk the Hilbert space is the span of all the vectors representing the $N \times N$ directed edges of the duplicated graph, i.e., $\mathcal{H} = \text{span}\lbrace\left|i\right>_1\left|j\right>_2,\ i,j = 0,1,...,N-1\rbrace = \mathbb{C}^N \otimes \mathbb{C}^N$, where the states with indexes $1$ and $2$ refer to the nodes on two copies of the original graph. The original formulation was based on two reflections, each around one of the two subspaces. However, there is an alternative formulation that generalizes the original one \cite{Notes} and it is more commonly used. We define the vectors:
\begin{equation}\label{psi_i}
	\left|\psi_i\right> := \left|i\right>_1 \otimes \sum_{k=0}^{N-1} \sqrt{G_{ki}}\left|k\right>_2,
\end{equation}
which are a superposition of the vectors representing the edges outgoing
from the $i^{th}$ vertex, whose coefficients are given by the square root of the $i^{th}$ column of the matrix $G$. From these vectors we can define a projector operator onto the subspace generated by them:
\begin{equation}
	\Pi := \sum_{i=0}^{N-1} \left|\psi_i\right>\left<\psi_i\right|.
\end{equation}
The quantum walk operator $U$ is defined as
\begin{equation}
	U := S(2\Pi - \mathbbm{1}),
\end{equation}
where $S$ is the swap operator between the two quantum registers, i.e.,
\begin{equation}
S := \sum_{i,j=0}^{N-1} \left|i,j\right>\left<j,i\right|.
\end{equation}
Let us define the adjacency matrix $A$ of a graph as a boolean matrix where $A_{ij} = 1$ if and only if there is an edge between nodes $i$ and $j$. When the transition matrix is obtained by normalizing the columns of the adjacency matrix the unitary operator $U$ corresponds to the one of the coined Grover quantum walk \cite{Grover,Notes}. Thus, the Szegedy's quantum walk can be understood as a coined quantum walk where the first register encodes the position in the graph, and the second register encodes the coin state. The original unitary operator formulated by Szegedy would be recovered taking the square of $U$, i.e. it would be $U^2$.

In order to formulate the semiclassical walk from Szegedy's quantum walk we can use the set of states $\left|\psi_i\right>$ as the proxy states in \eqref{proxies}. An example of implementation of a semiclassical Szegedy's walk is shown in Figure \ref{F:Szegedy_sc_1}. Let us denote $x_{t_c}$ as the position at classical time step $t_c$. Thus, we start at node $x_0$. We prepare the proxy $\left|\psi_{x_0}\right>$ and perform the quantum evolution, parameterized by the quantum time $t_q$. After measuring the first register the system collapses to a particular node $x_1$ in the first register, and we do not worry about the state of the second register. We use the measured information about the node to reset the system and prepare the new proxy for the node we have just measured, $\left|\psi_{x_1}\right>$. This process is then repeated the number of classical steps as desired. We call this a semiclassical walk of class I since we are measuring in the first register

Although it is common to measure the first register, there are other applications where the second register is measured instead to obtain the information about the nodes. An example is the quantum PageRank algorithm \cite{Paparo1,Paparo2,APR}. Thus, we can define a semiclassical walk of class II by measuring in the second register. An implementation is shown in Figure \ref{F:Szegedy_sc_2}. In this case the information of the nodes are obtained measuring the second register, but the proxies are prepared in the same form as before, as per equation \ref{psi_i}.

Finally, we can simulate both classes of semiclassical walks as classical walks with a semiclassical transition matrix. Let us use a left-subscript in the semiclassical matrix to denote the class of the walk. Then, the semiclassical matrices are obtained as:
\begin{equation}\label{G1}
\tensor[_1]{G}{}^{(t_q)}_{ji} := \left|\left|\tensor[_1]{\left<j\right|U^{t_q}\left|\psi_i\right>}{}\right|\right|^2,
\end{equation}
\begin{equation}\label{G2}
\tensor[_2]{G}{}^{(t_q)}_{ji} := \left|\left|\tensor[_2]{\left<j\right|U^{t_q}\left|\psi_i\right>}{}\right|\right|^2.
\end{equation}

From the semiclassical matrices we can formulate some theorems about the semiclassical Szegedy's walk.

\textbf{Theorem 1: Classical limit I.} The classical walk is recovered for the semiclassical walk of class I with a quantum time $t_q=1$, thus:
\begin{equation}
\tensor[_1]{G}{}^{(1)} = G.
\end{equation}
\textbf{Proof:} We start calculating the quantum state that results of applying the unitary evolution once.
\begin{equation}
U\left|\psi_i\right> = S(2\Pi - \mathbbm{1})\left|\psi_i\right> = S \left|\psi_i\right>,
\end{equation}
since $\Pi\left|\psi_i\right> = \left|\psi_i\right>$, due that the space where $\Pi$ projects is the subspace spanned by the states $\left|\psi_i\right>$. The swap operator swaps the states between both registers, so
\begin{equation}\label{Upsi}
	U\left|\psi_i\right> = \sum_{k=0}^{N-1} \sqrt{G_{ki}} \left|k\right>_1 \left|i\right>_2.
\end{equation}
To obtain the semiclassical matrix $_1G^{(1)}$ we take the inner product with the computational basis of the first register and take the squared modulus.
\begin{equation}
\tensor[_1]{\left<j\right|U\left|\psi_i\right>}{} = \sum_{k=0}^{N-1}\left(\delta_{jk}\sqrt{G_{ki}}\left|i\right>_2\right)= \sqrt{G_{ji}}\left|i\right>_2,
\end{equation}
\begin{equation}
\tensor[_1]{G}{}^{(1)}_{ji} = \left|\left|\tensor[_1]{\left<j\right|U\left|\psi_i\right>}{}\right|\right|^2 = G_{ji}. \qed
\end{equation}
This theorem reinforces the idea that the set of states $\left|\psi_i\right>$ is natural as proxies for the semiclassical walks, since the classical walk is obtained in the limit of only applying once the unitary evolution between measurements, which corresponds to a lack of a coherent quantum evolution.

\textbf{Theorem 2: Classical limit II.} The classical walk is recovered for the semiclassical walk of class II with a quantum time $t_q=2$, thus:
\begin{equation}\label{Th2}
\tensor[_2]{G}{}^{(2)} = G.
\end{equation}
For a proof see Supplementary Material (SM) \cite{SM}. This theorem makes us think that when measuring the second register it is more natural to use the Szegedy's original quantum unitary evolution $U^2$ instead of $U$. This is indeed what is done in the quantum PageRank \cite{Paparo1,Paparo2,APR}.

\textbf{Theorem 3: Equivalence between semiclassical classes.} The semiclassical walk of class I obtained with a quantum time $t_q$ is the same as the one of class II obtained with a quantum time $t_q+1$:
\begin{equation}\label{Th3}
\tensor[_1]{G}{}^{(t_q)} = \tensor[_2]{G}{}^{(t_q+1)}.
\end{equation}
For a proof see SM \cite{SM}. Due to this theorem, in the following we will only regard to the semiclassical walks of class I, since we are dealing with the general single operator $U$ in this work. However, for other scenarios where the evolution were performed with $U^2$, both classes would not be equivalent. There would not be equivalence either if the operator were modified with oracles, as done for quantum search \cite{Searchrank,S_queries}, or modified with arbitrary phase rotations \cite{APR}.

\section{Szegedy Semiclassical Walk on 1D Cycles}\label{Cycles}

\begin{figure*}
	\centering
	\subfigure[]{\includegraphics[scale=0.375]{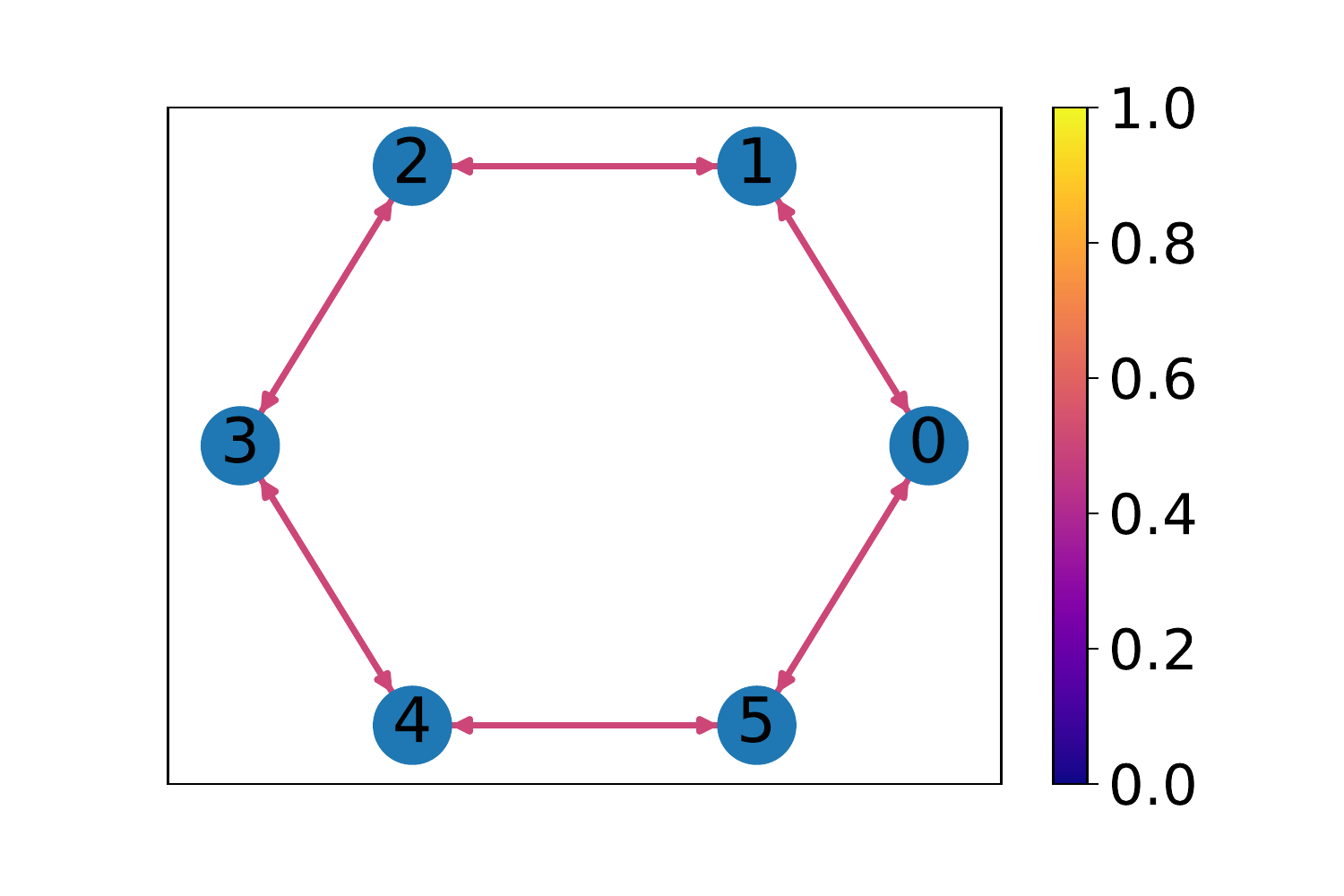}}
	\subfigure[]{\includegraphics[scale=0.375]{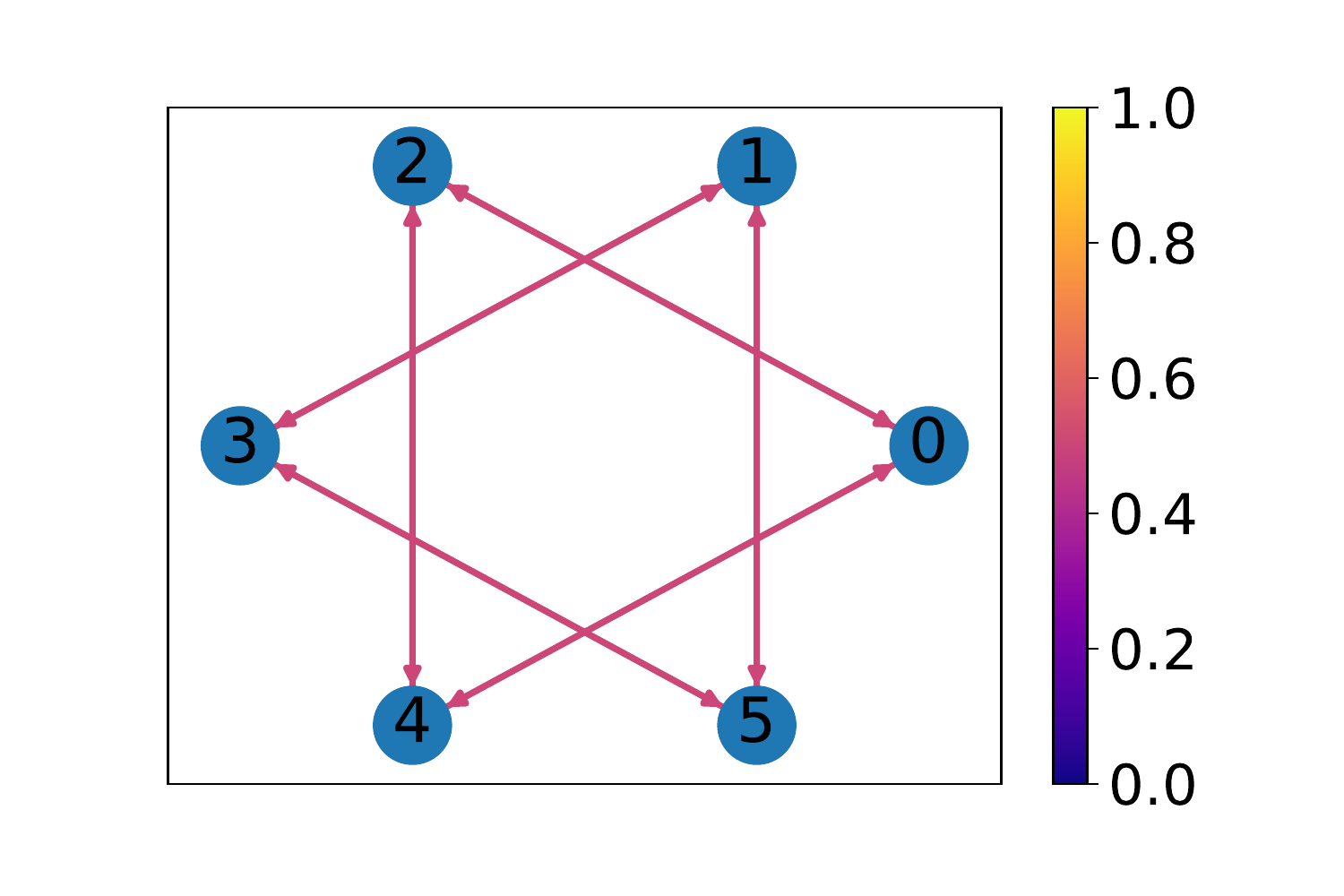}}
	\subfigure[]{\includegraphics[scale=0.375]{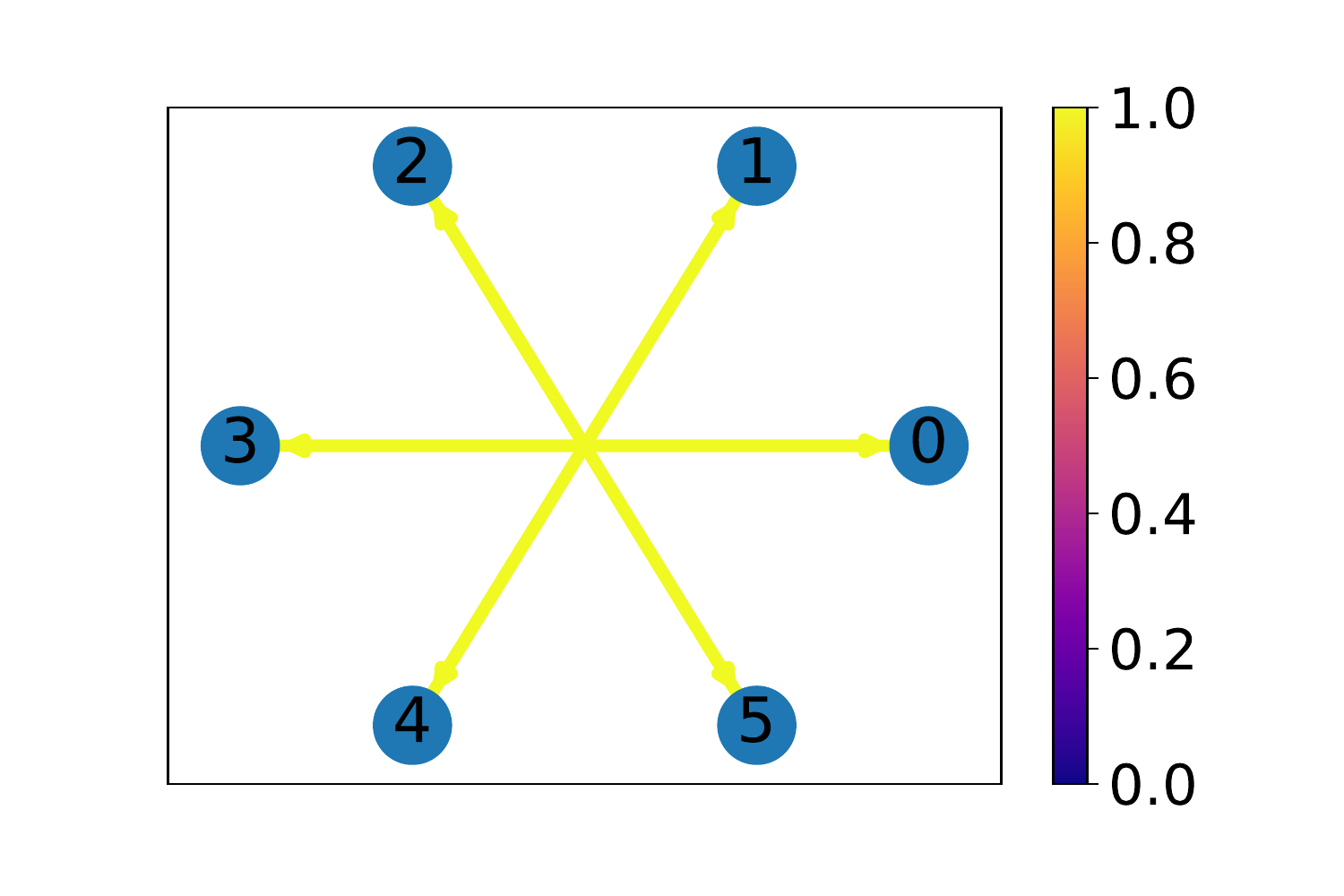}}
	\subfigure[]{\includegraphics[scale=0.375]{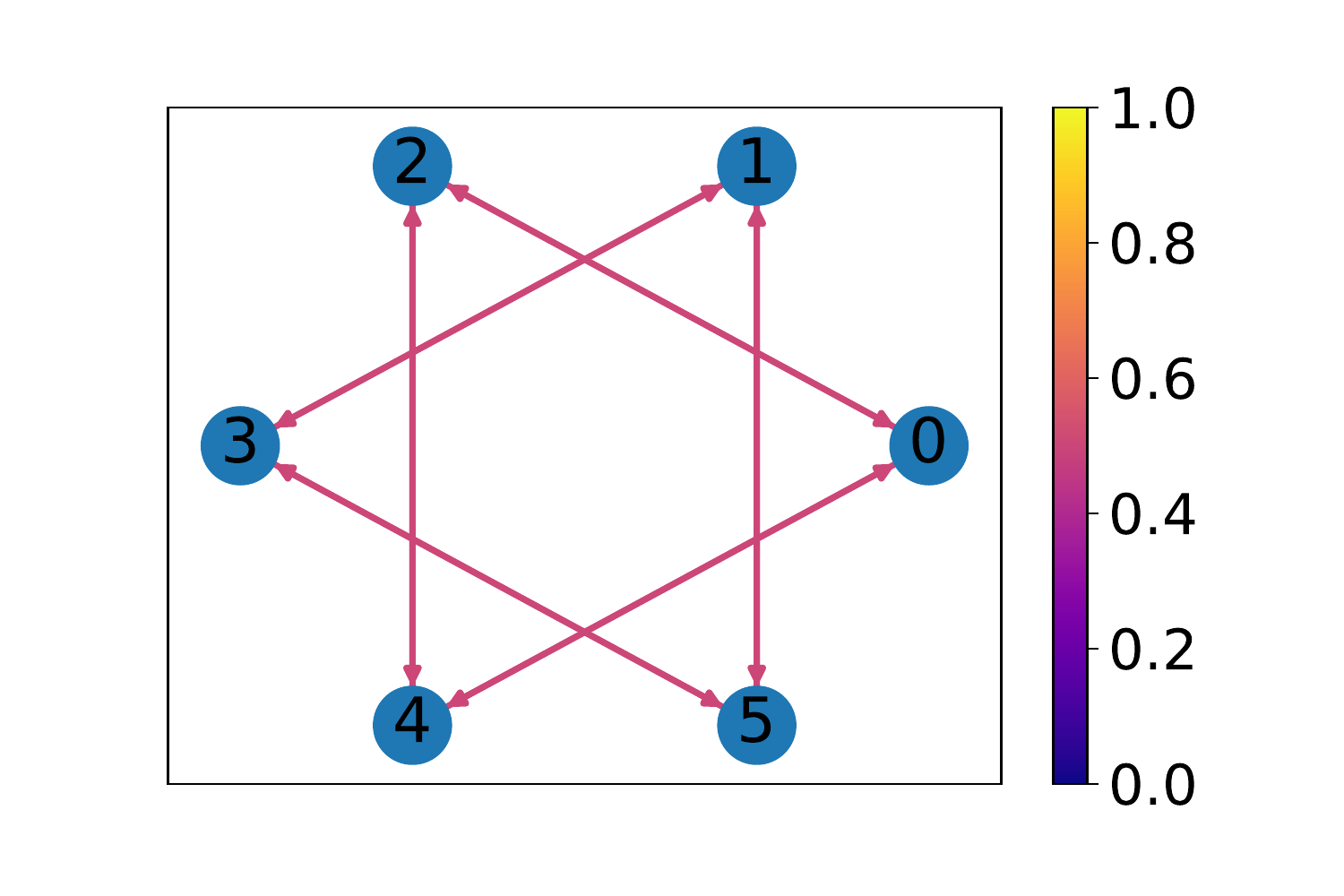}}
	\subfigure[]{\includegraphics[scale=0.375]{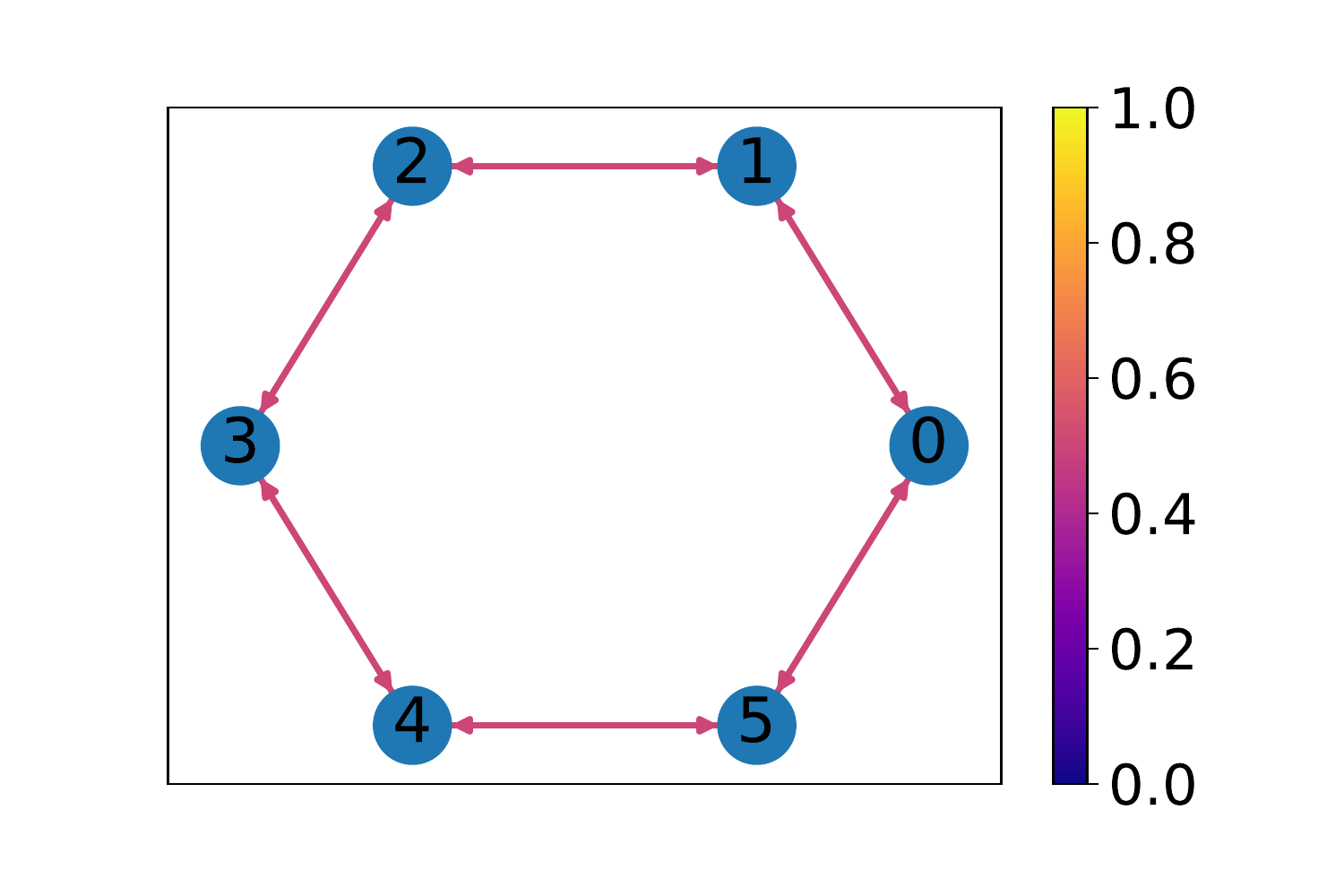}}
	\subfigure[]{\includegraphics[scale=0.375]{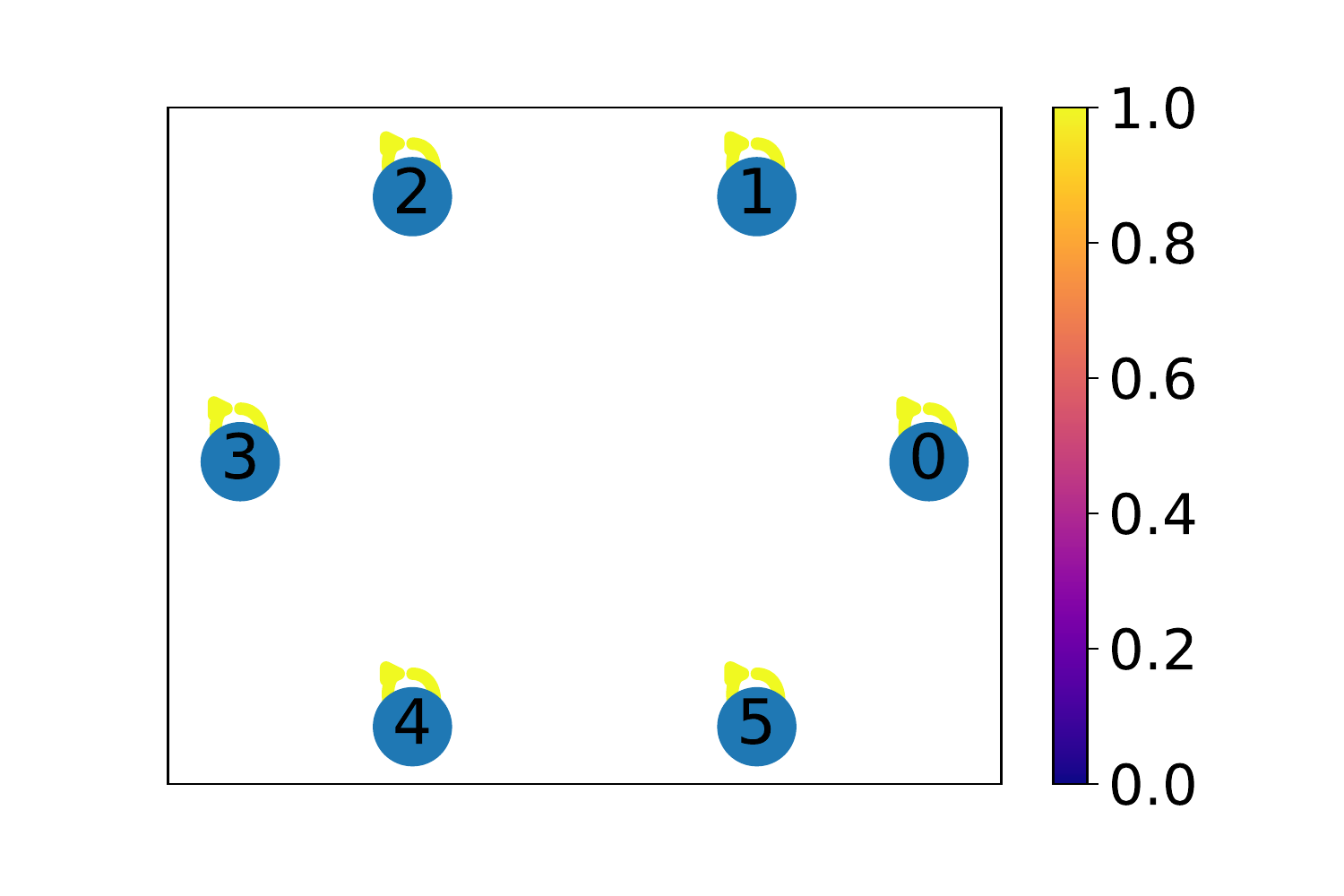}}
	\caption{Semiclassical graphs for the cycle with $N=6$ nodes for a) $t_q=1$, b) $t_q=2$, c) $t_q=3$, d) $t_q=4$, e) $t_q=5$, f) $t_q=6$. The weights of the edges are represented by the colormap. In this case there are only two possible weights: $0.5$ represented by a magenta line, or $1$ represented by a yellow line. All the edges are bidirectional. The graphs have been plotted using the python library NetworkX \cite{NetworkX}.}
	\label{F:N=6}
\end{figure*}

Once we have defined the semiclassical Szegedy's walk, let us see some examples in 1D lattices. We are going to analytically solve the problem for the infinite line, and after that put cyclic boundary conditions to obtain the semiclassical walks on 1D cycles.

The classical matrix of the walk is given by
\begin{equation}
G_{ji} = \frac{1}{2}\delta_{i,j+1} + \frac{1}{2}\delta_{i,j-1},
\end{equation}
so that a walker at node $i$ can jump to either node $i+1$ or $i-1$ with a $50 \%$ of probability for both cases. The proxy states in the Szegedy's semiclassical walk are
\begin{equation}\label{psi_par}
\left|\psi_i\right> = \left|i\right>_1 \otimes \frac{1}{\sqrt{2}}\left(\left|i-1\right>_2 + \left|i+1\right>_2\right).
\end{equation}
We can define a set of orthogonal states to these proxy states as follows:
\begin{equation}\label{psi_perp}
\left|\psi^\perp_i\right> := \left|i\right>_1 \otimes \frac{1}{\sqrt{2}}\left(\left|i-1\right>_2 - \left|i+1\right>_2\right).
\end{equation}
With these sets we can calculate easily the action of the unitary operator $U$ over the set of states $\left|i\right>_1\left|i \pm 1\right>_2$. These states can be expressed as
\begin{equation}
\left|i\right>_1\left|i-1\right>_2 = \frac{1}{\sqrt{2}}\left(\left|\psi_i\right>+\left|\psi^\perp_i\right>\right),
\end{equation}
\begin{equation}
\left|i\right>_1\left|i+1\right>_2 = \frac{1}{\sqrt{2}}\left(\left|\psi_i\right>-\left|\psi^\perp_i\right>\right).
\end{equation}
Since the states $\left|\psi^\perp_i\right>$ are perpendicular to all states $\left|\psi_i\right>$, they are in the kernel of the projector, so $\Pi\left|\psi_i^\perp\right> = 0$. Using the expressions \eqref{psi_par} and \eqref{psi_perp}, the reflection part of the unitary operator yields
\begin{equation}
(2\Pi - \mathbbm{1})\left|i\right>_1\left|i-1\right>_2 = \left|i\right>_1\left|i+1\right>_2,
\end{equation}
\begin{equation}
(2\Pi - \mathbbm{1})\left|i\right>_1\left|i+1\right>_2 = \left|i\right>_1\left|i-1\right>_2.
\end{equation}
Finally, we apply the swap between the two registers, obtaining the action of $U$:
\begin{equation}\label{i-}
	U\left|i\right>_1\left|i-1\right>_2 = \left|i+1\right>_1\left|i\right>_2,
\end{equation}
\begin{equation}\label{i+}
	U\left|i\right>_1\left|i+1\right>_2 = \left|i-1\right>_1\left|i\right>_2.
\end{equation}
With \eqref{i-} and \eqref{i+} we can calculate the quantum evolution of the proxy states:
\begin{equation}
	U\left|\psi_i\right> =\frac{1}{\sqrt{2}}\left(\left|i+1\right>_1\left|i\right>_2 + \left|i-1\right>_1\left|i\right>_2\right),
\end{equation}
and for a general number $t_q$ of quantum steps:
\begin{eqnarray}\label{SQW}
	U^{t_q}\left|\psi_i\right> &=&\frac{1}{\sqrt{2}}\left(\left|i+t_q\right>_1\left|i+t_q-1\right>_2 \right.\nonumber\\
	& &+\left. \left|i-t_q\right>_1\left|i-t_q+1\right>_2\right).
\end{eqnarray}
We see then that the quantum walk starting from a single node moves apart from that node in a symmetric form. The walker jumps $t_q$ times from the starting node $i$, reaching nodes $\pm t_q$ with a probability of $50 \%$.

If we impose cyclic boundary conditions, then we have the identification $-N=0=N$, and the same for each two integers with a difference of $N$. Then, there are two interesting cases. The first one is when $N$ is even and $t_q = N/2$. For the sake of simplicity, let us see the effect over the state $\left|\psi_0\right>$ in a graph with $N=6$, so $t_q=3$ and
\begin{eqnarray}
U^3\left|\psi_0\right> &=&\frac{1}{\sqrt{2}}\left(\left|3\right>_1\left|2\right>_2 + \left|-3\right>_1\left|-2\right>_2\right)\nonumber\\ &=&\frac{1}{\sqrt{2}}\left(\left|3\right>_1\left|2\right>_2 + \left|3\right>_1\left|4\right>_2\right) = \left|\psi_3\right>,
\end{eqnarray}
where we have used the boundary conditions to identify $-3$ with $3$ and $-2$ with $4$. In this case the walker reach the same node from both sides, so the probability of measuring it is of $100 \%$. The second case is when $t_q = N$ for any value of $N$. In that case:
\begin{eqnarray}
U^N\left|\psi_0\right> &=&\frac{1}{\sqrt{2}}\left(\left|N\right>_1\left|N-1\right>_2 + \left|-N\right>_1\left|-N+1\right>_2\right)\nonumber\\
&=&\frac{1}{\sqrt{2}}\left(\left|0\right>_1\left|-1\right>_2 + \left|0\right>_1\left|1\right>_2\right) = \left|\psi_0\right>,
\end{eqnarray}
so the walker start at the same point with certainty. Thus, the Szegedy's quantum walk has a period of $N$ over the proxy states $\left|\psi_i\right>$. This is a great difference with the Hadamard coined quantum walk in 1D cycles, which is only periodic for a few values of $N$ \cite{Portugal}.

\begin{figure}[htpb]
	\centering
	\includegraphics[scale=0.5]{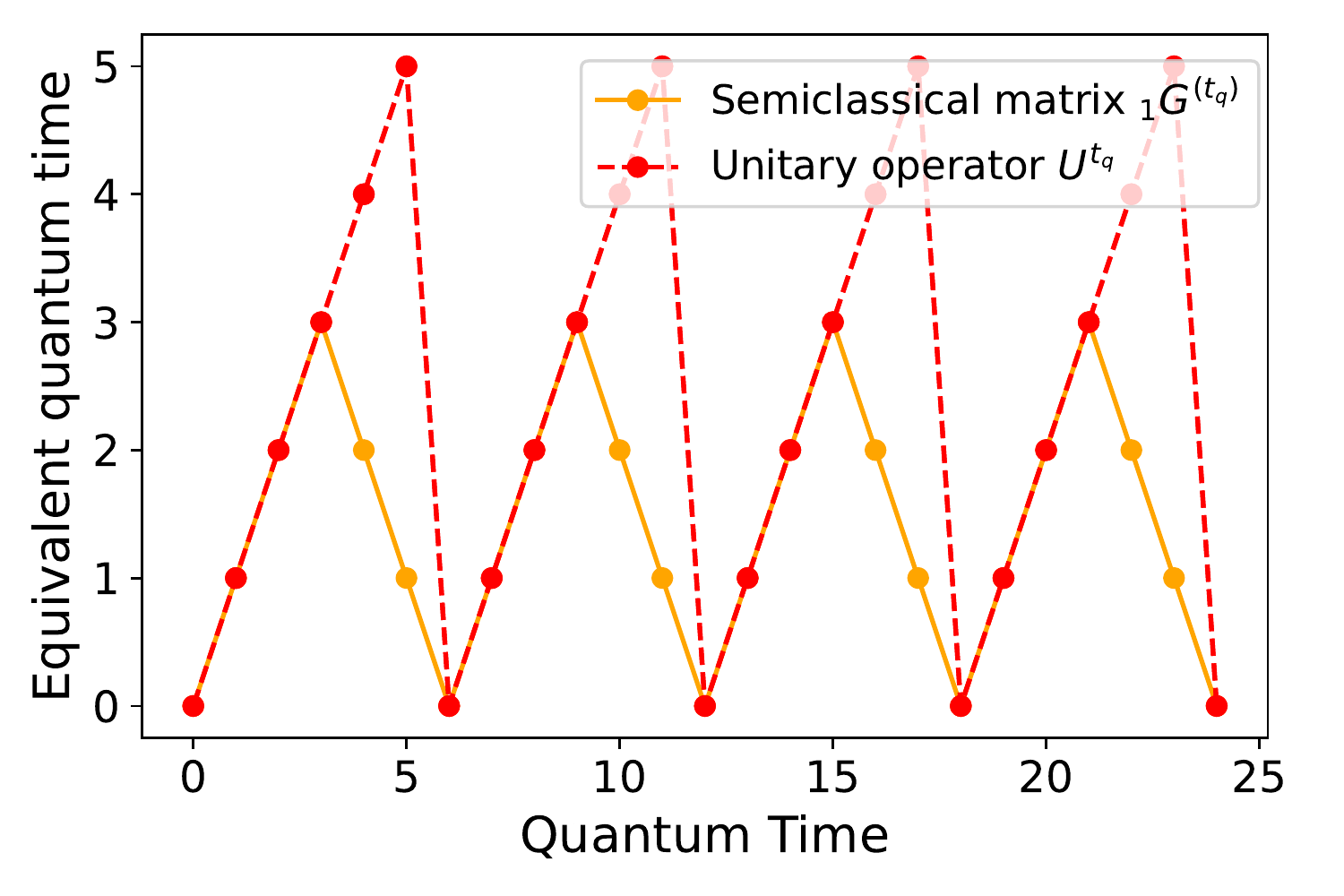}
	\caption{Periodicity of the semiclassical matrices and the unitary evolution for the cycle with $N=6$ nodes. At each quantum time it is represented the minimum value of $t_q$ for which the matrix is equal. For example, for $t_q=4$ the semiclassical matrix is equal to the one at $t_q=2$. However, the unitary operator is not still repeated, so that the equivalent quantum time is also $t_q=4$. Time $t_q=0$ is not an actual walk, but is used to represent that the matrix is equal to the identity.}
	\label{F:N=6_period}
\end{figure}

Finally, from \eqref{SQW} we can calculate the semiclassical matrices using \eqref{G1} for the semiclassical family of class I:
\begin{equation}
\tensor[_1]{G}{}^{(t_q)}_{ji} = \frac{1}{2}\left(\delta_{i,j+t_q} + \delta_{i,j-t_q}\right).
\end{equation}
This is equivalent to a classical walk where each node $i$ connects only to nodes $i\pm t_q$. Due to the periodicity of $U$ over the proxy states, the semiclassical family will also have a periodicity in the quantum time, thus
\begin{equation}
\tensor[_1]{G}{}^{(t_q)} = \tensor[_1]{G}{}^{(t_q+N)},
\end{equation}
so there will be at most $N$ different semiclassical walks.

\begin{figure*}
	\centering
	\subfigure[]{\includegraphics[scale=0.375]{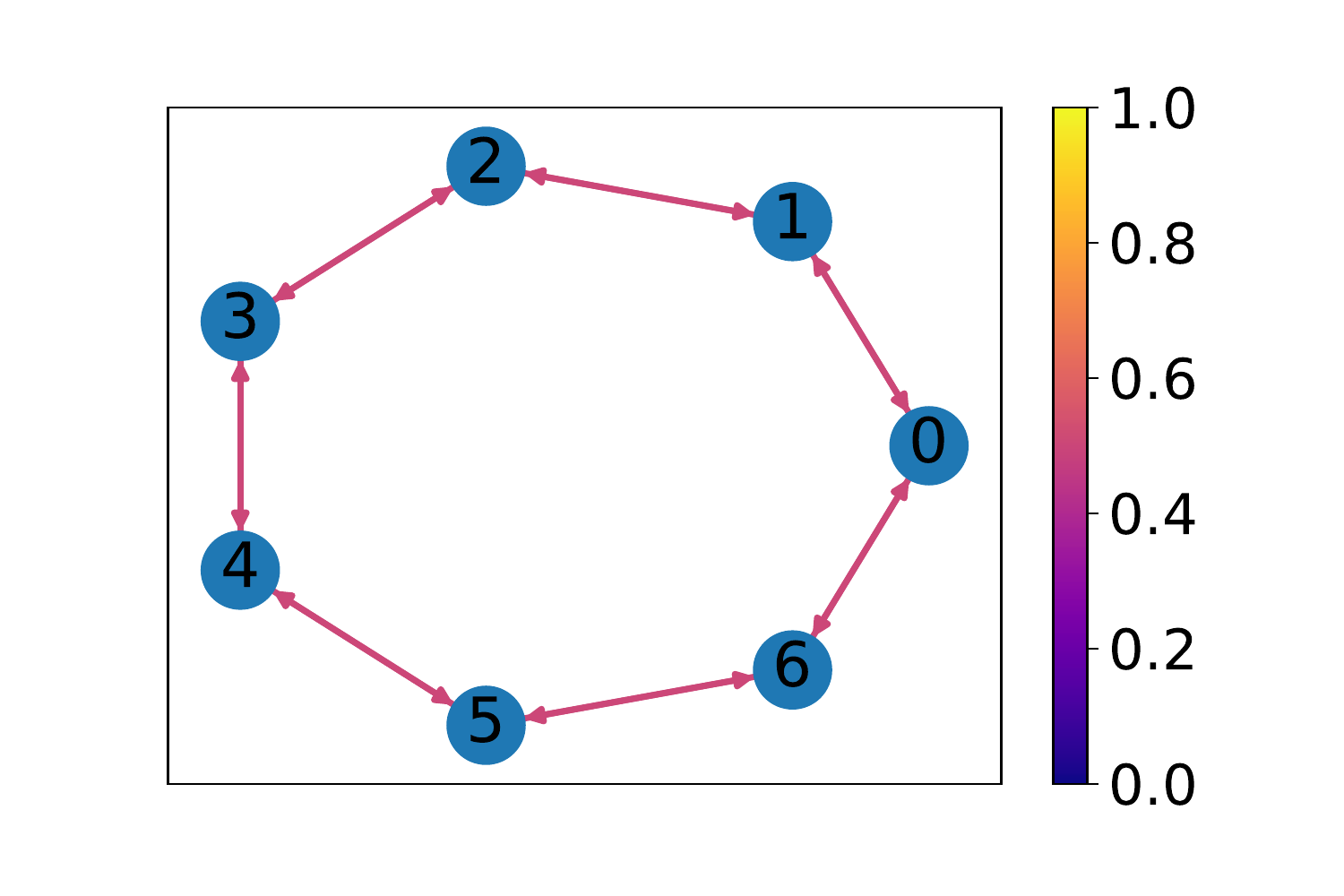}}
	\subfigure[]{\includegraphics[scale=0.375]{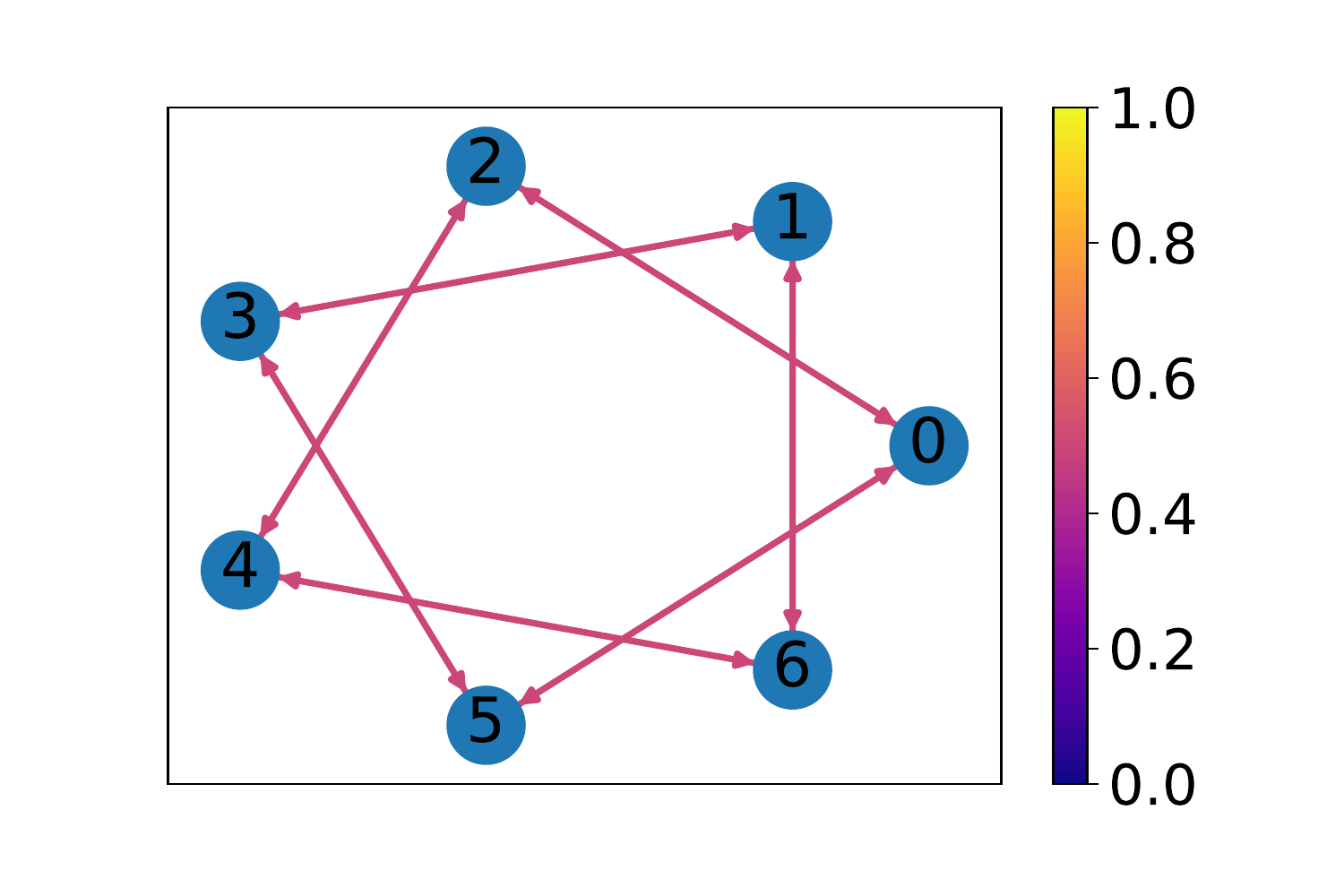}}
	\subfigure[]{\includegraphics[scale=0.375]{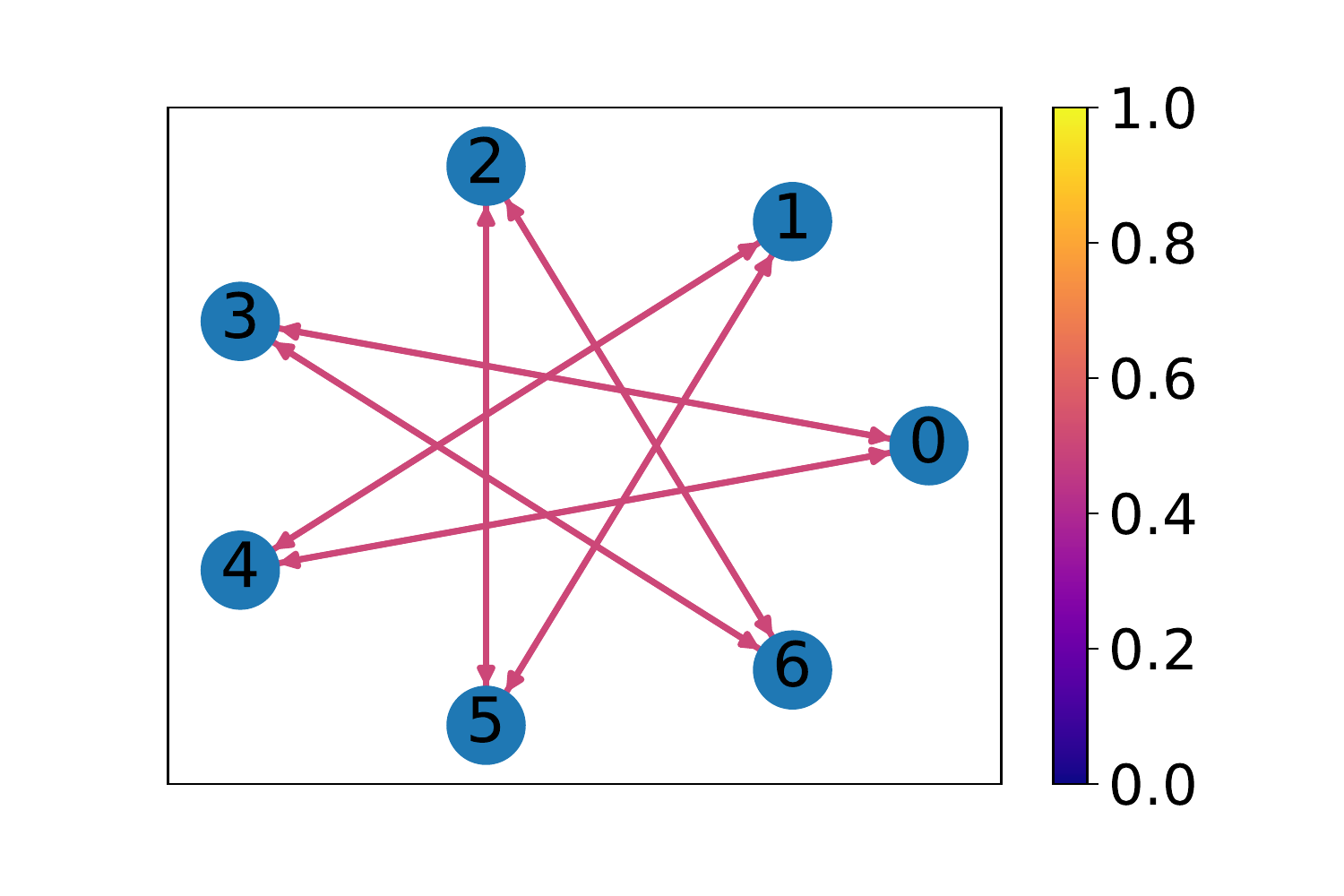}}
	\subfigure[]{\includegraphics[scale=0.375]{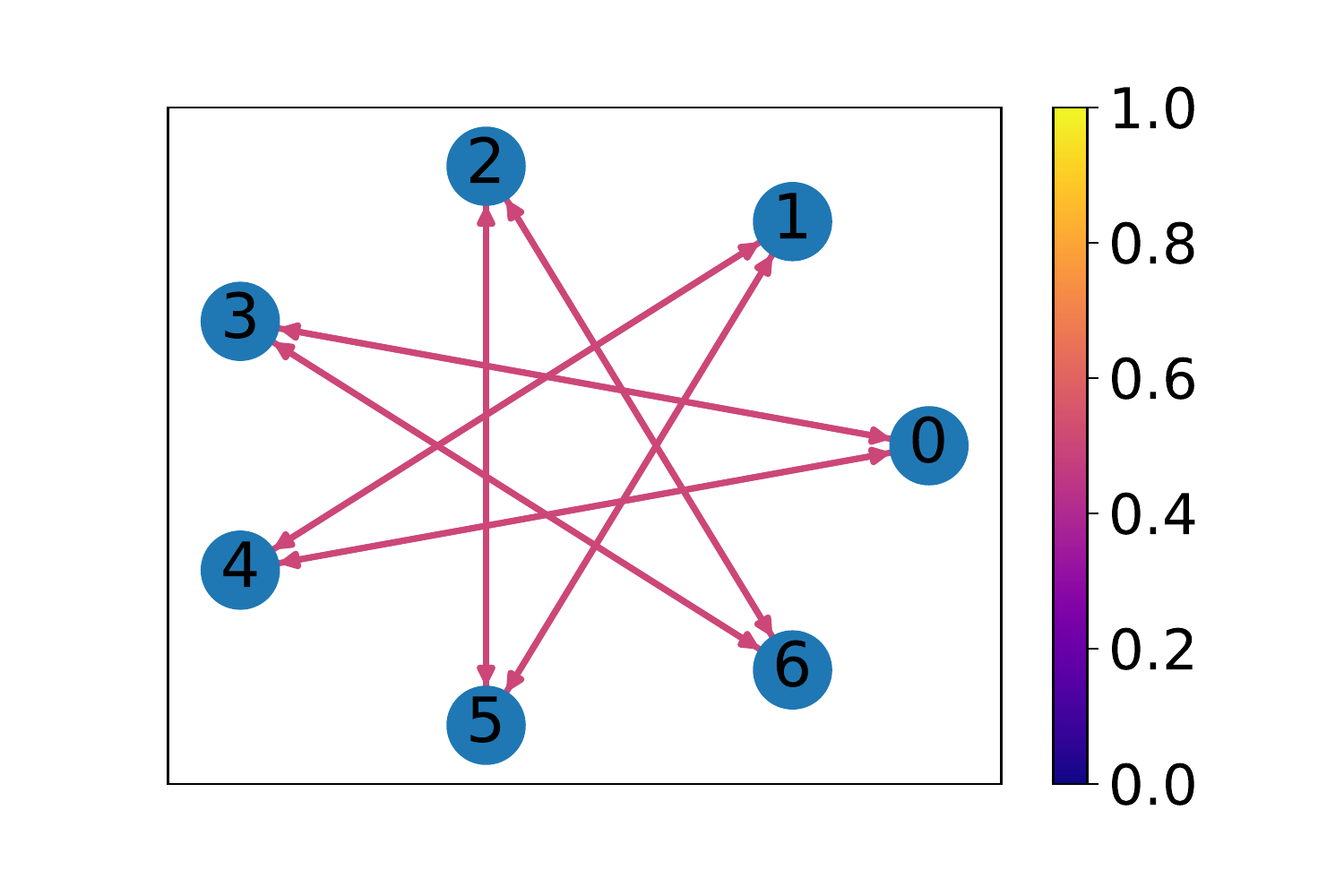}}
	\subfigure[]{\includegraphics[scale=0.375]{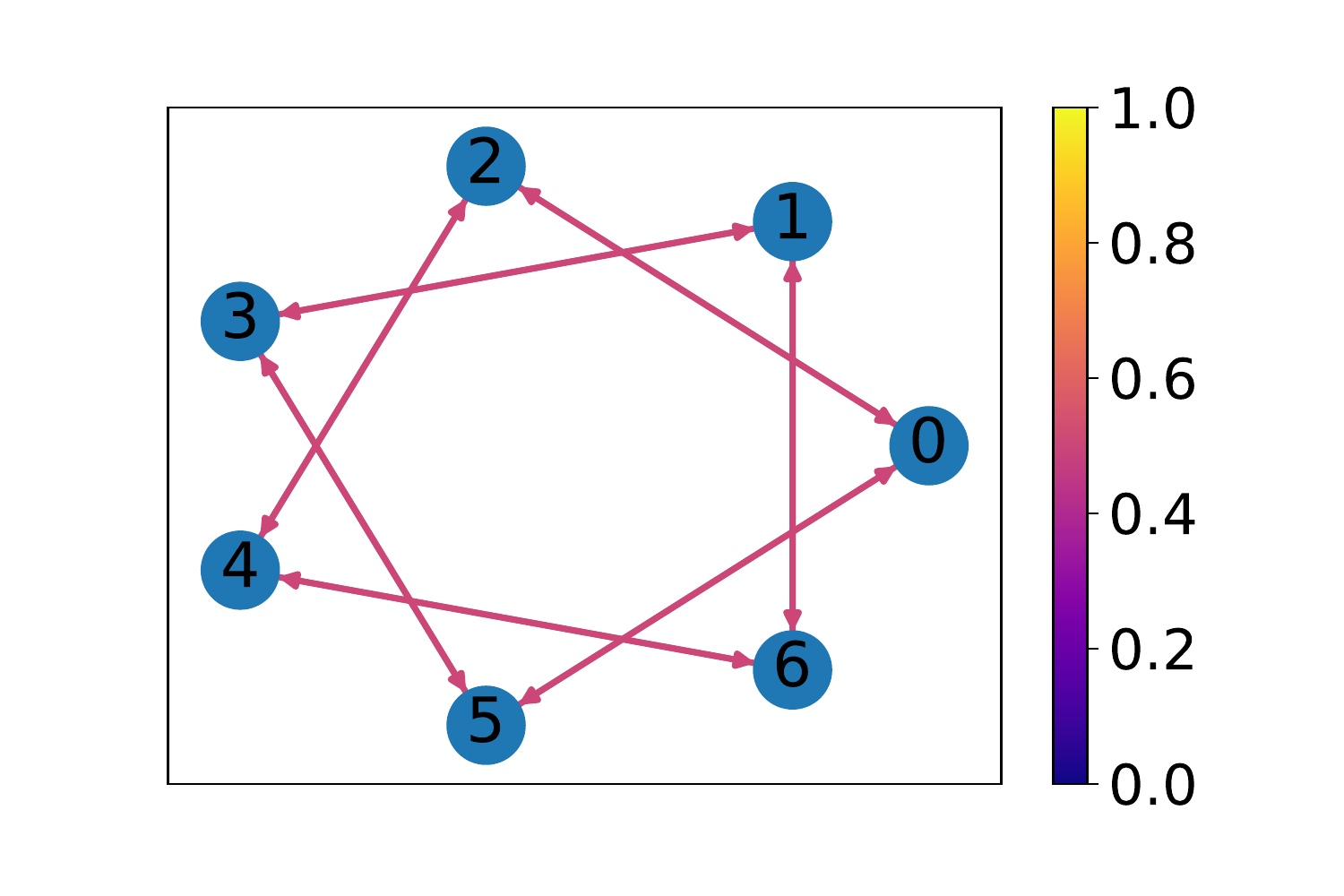}}
	\subfigure[]{\includegraphics[scale=0.375]{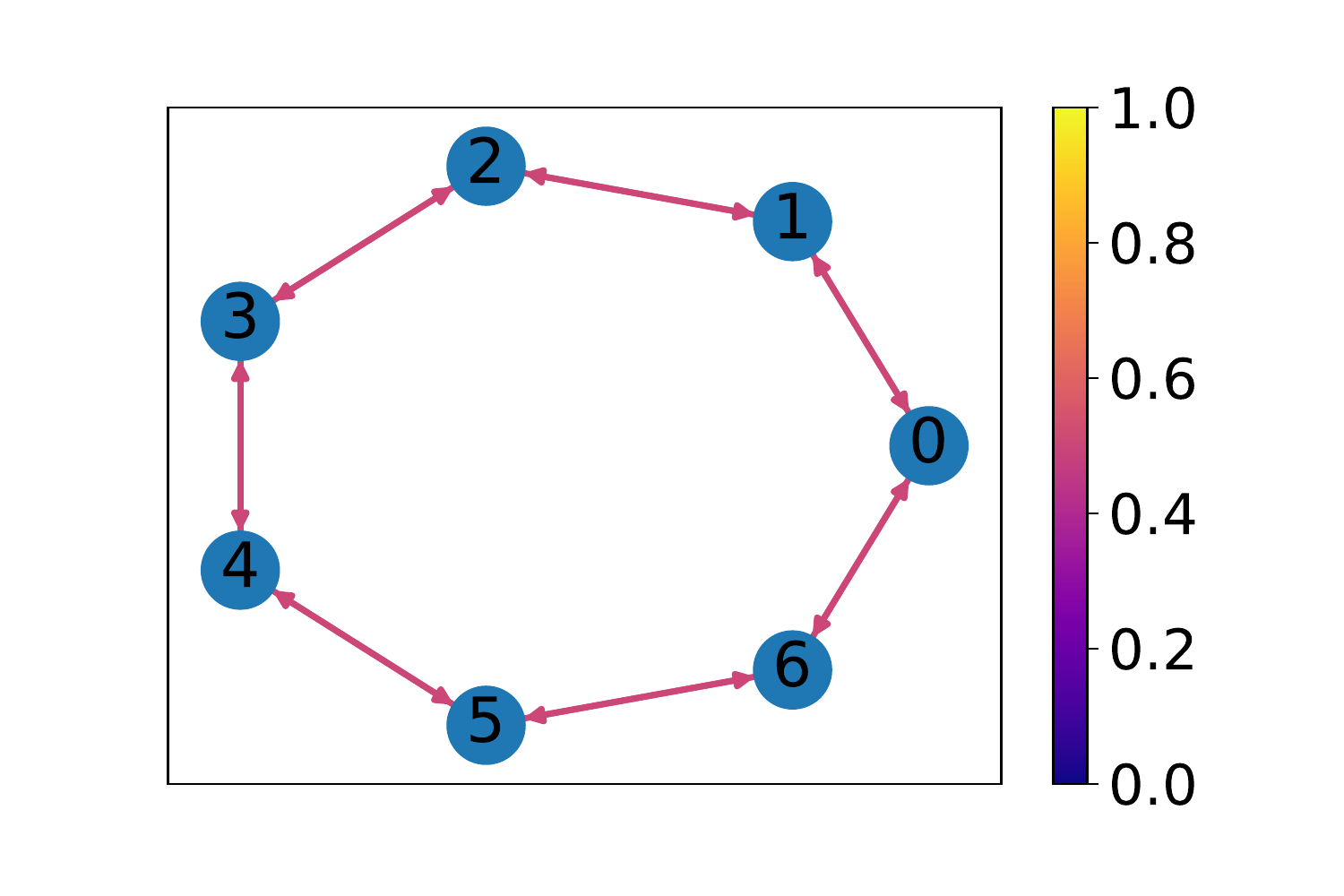}}
	\subfigure[]{\includegraphics[scale=0.375]{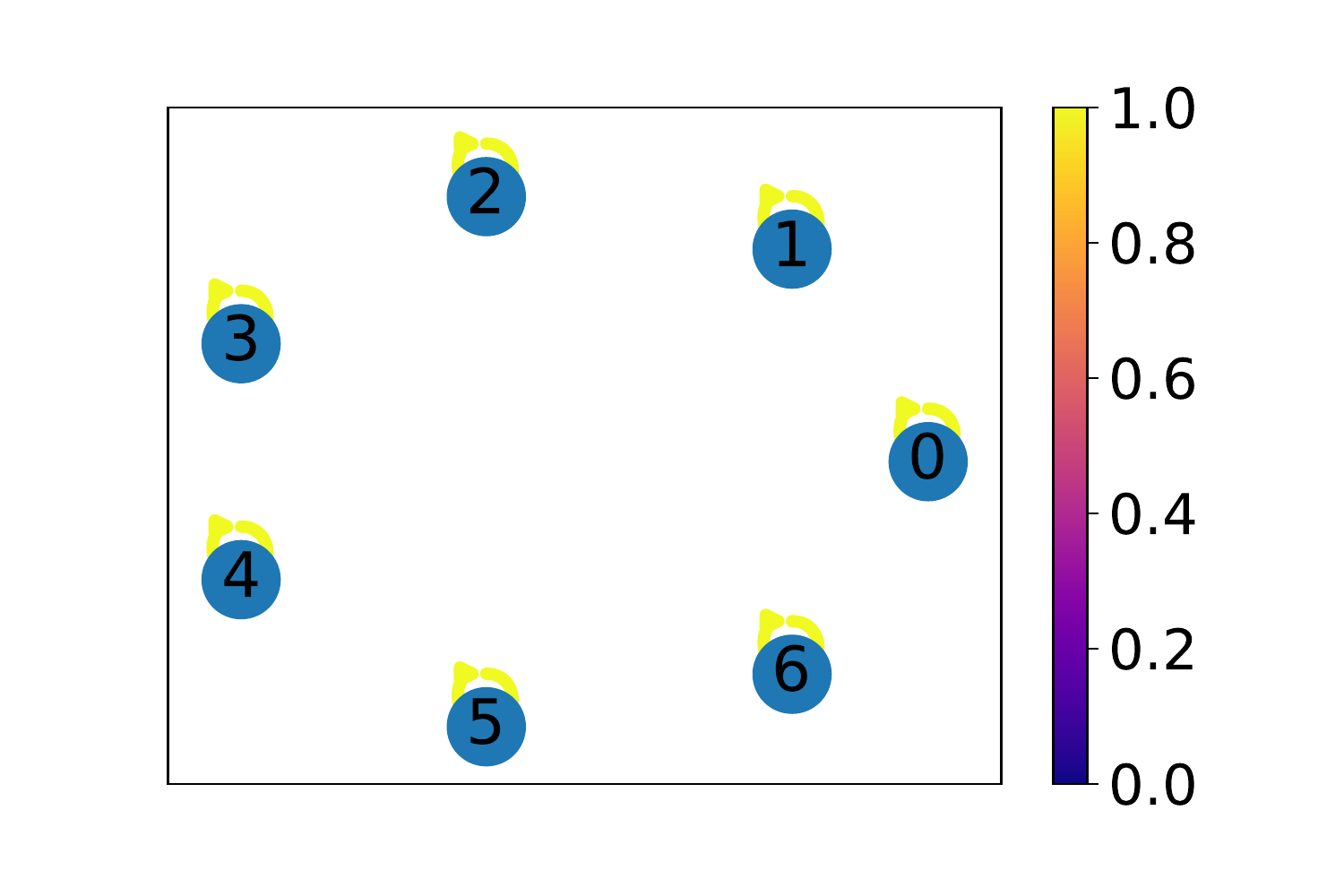}}
	\subfigure[]{\includegraphics[scale=0.375]{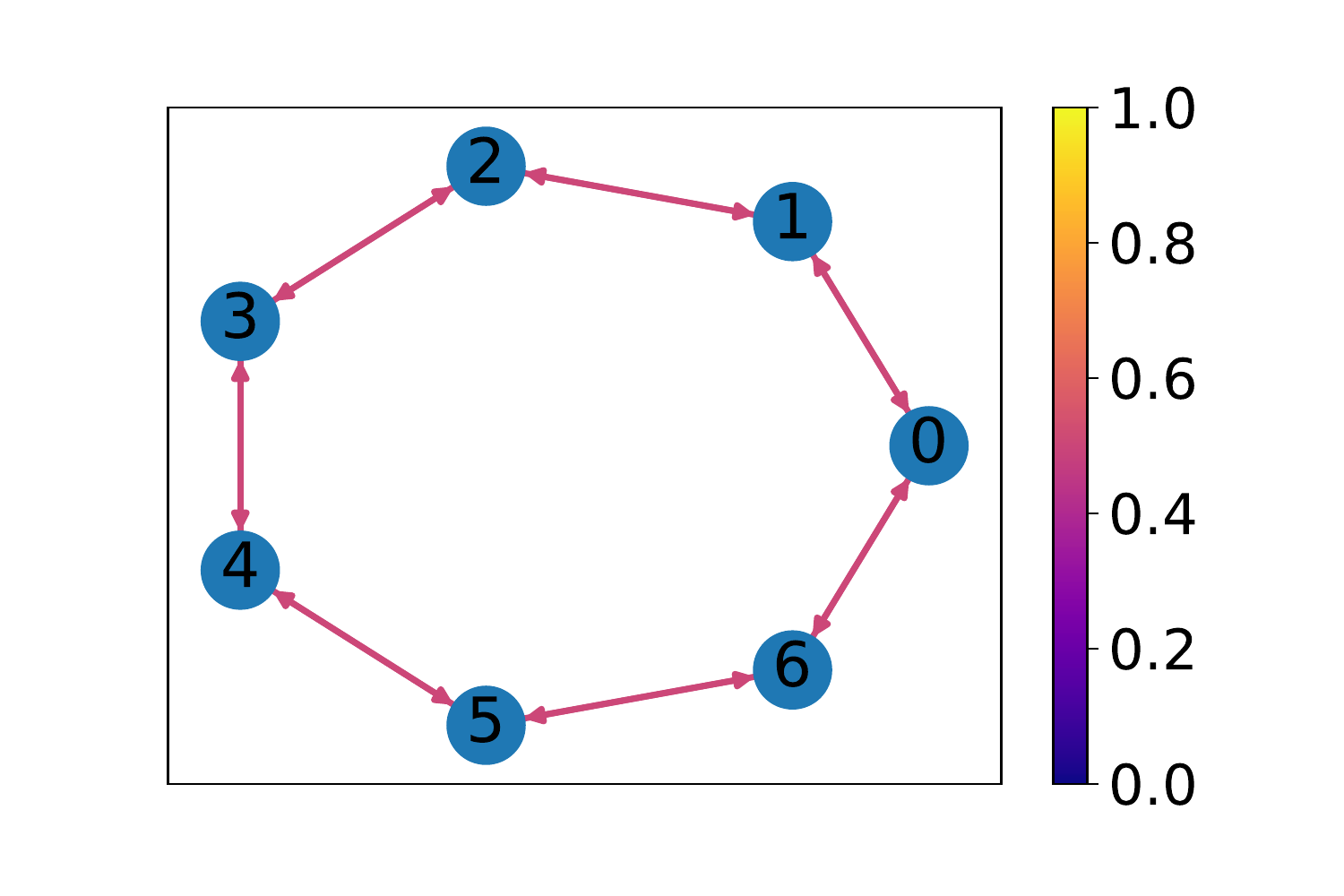}}
	\subfigure[]{\includegraphics[scale=0.375]{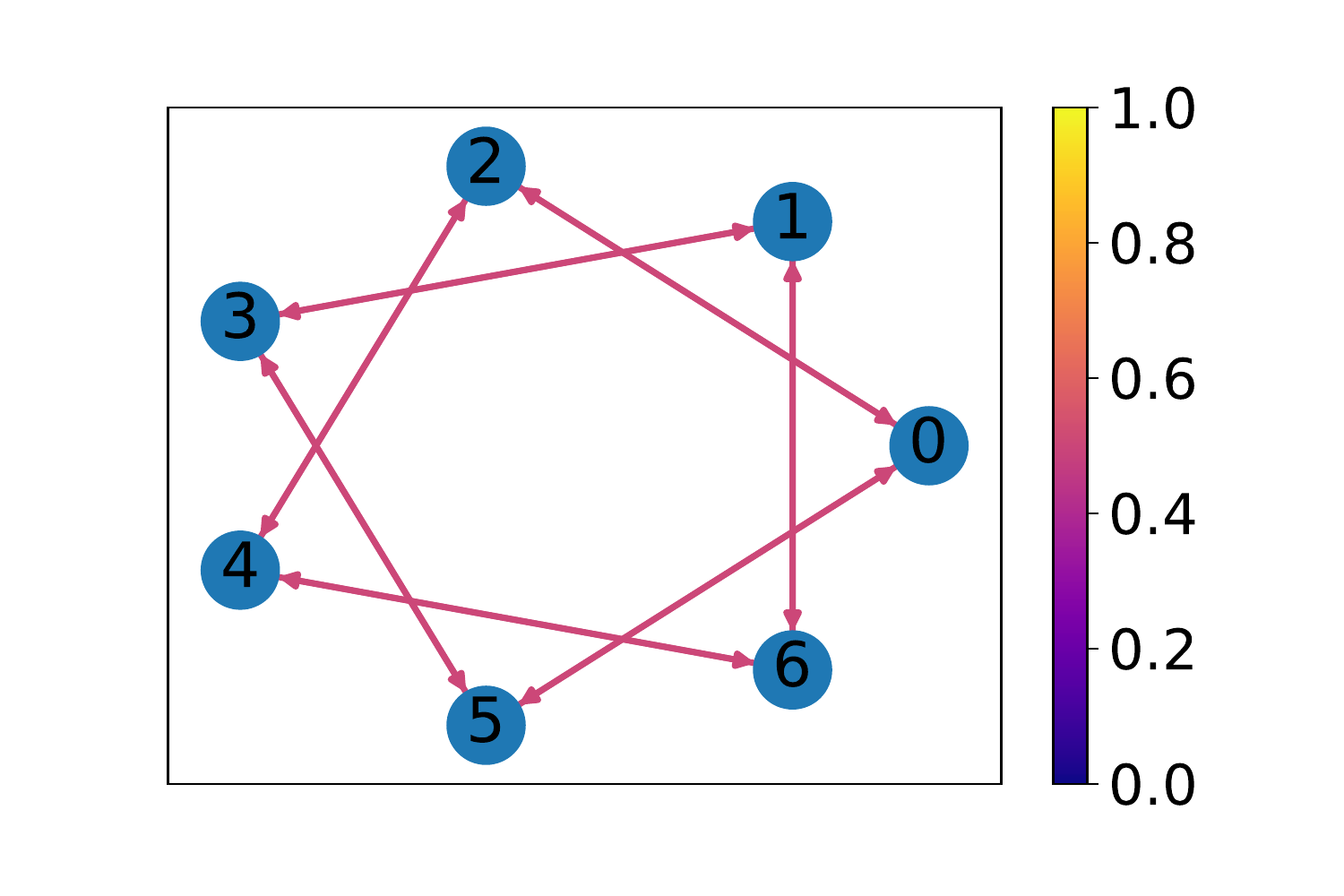}}
	\caption{Semiclassical graphs for the cycle with $N=7$ nodes for a) $t_q=1$, b) $t_q=2$, c) $t_q=3$, d) $t_q=4$, e) $t_q=5$, f) $t_q=6$, g) $t_q=7$, h) $t_q=8$, i) $t_q=9$. The weights of the edges are represented by the colormap. In this case there are only two possible weights: $0.5$ represented by a magenta line, or $1$ represented by a yellow line. All the edges are bidirectional. The graphs have been plotted using the python library NetworkX \cite{NetworkX}.}
	\label{F:N=7}
\end{figure*}

Since a classical walk occurs in a graph where the edges encode the probabilities of the walker jumping from one node to another, we can also represent the semiclassical walk family as a set of weighted graphs, denoted as semiclassical graphs. In Figure \ref{F:N=6} it is shown an example of the semiclassical graphs for the cycle with $N=6$ nodes. For the first quantum time, $t_q=1$, we obtain the same graph than in the classical walk, so each node links to its immediate neighbors. The same result is obtained for $t_q=5$. However, for other values of the quantum time we obtain genuine walks. For $t_q=2$ and $t_q=4$ each node connects to the second nearest neighbors, so the graphs breaks into triangles. Thus, if a particle start at node $0$ it will perform a walk equivalent to a classical one in the triangle formed by nodes $0$, $2$ and $4$. For $t_q =3$ each node links only to the opposite node in the graph, breaking the graph into lines of 2 nodes. Finally, for $t_q=6$ each node links only to itself with a loop, so the graphs breaks into single nodes. This would be equivalent to $t_q=0$, which is not actually a walk since there is not quantum evolution, so that the transition matrix is just the identity. For a larger value of the quantum time the sequence of graphs is repeated due to the periodicity of $N=6$. The breaking in the connectivity of the graph is due to a degeneration of the eigenvalue $1$ of the semiclassical matrix, which agrees with the results of the continuous quantum time version in \cite{MIQW} where the same cycle with six nodes was also broken for concrete values of the quantum time.

We have seen that a priori we could have $N$ different semiclassical graphs. Nevertheless, here there are some graphs inside a period that are repeated. This is due to the fact that the semiclassical walks comes from the projection of quantum states so that different quantum states can yield the same position after measurement. As an example, note that the quantum states after the evolution from $\left|\psi_0\right>$ are not the same for $t_q=2$ and $t_q=4$:
\begin{equation}
U^2\left|\psi_0\right> =\frac{1}{\sqrt{2}}\left(\left|2\right>_1\left|1\right>_2 + \left|4\right>_1\left|5\right>_2\right),
\end{equation}
\begin{equation}
U^4\left|\psi_0\right> =\frac{1}{\sqrt{2}}\left(\left|4\right>_1\left|3\right>_2 + \left|2\right>_1\left|3\right>_2\right).
\end{equation}
Nevertheless, when the first register is measured, in both cases there is a $50 \%$ of probability for measuring node $2$ or node $4$, giving rise to the same semiclassical walk. To see better how the semiclassical matrices between different quantum times are related, in Figure \ref{F:N=6_period} it is shown the oscillation of the semiclassical matrices $_1G^{(t_q)}$ with respect to the quantum time, where we can see that there are only 4 different semiclassical walks in the family. In the same figure it is shown the period of the unitary operator $U$, so we effectively see that there are six different operators $U^{t_q}$.

The number of different semiclassical walks depend on how many jumps the walker makes between the measurements. The number of jumps is just $t_q$, and since it jumps in both directions, due to the cyclic boundary conditions we would have in general that the number of different graphs is
\begin{equation}\label{n_graphs}
\# \ \text{graphs} = \lfloor N/2 \rfloor +1.
\end{equation}
Furthermore, the type of subgraphs that the classical graph can be broken into depends on how the number of nodes $N$ can be factorized. For $N=6$ we have $6 = 2\times 3 = 1\times 6$, so we can have one hexagon, two triangles, three lines or six separate nodes.

As another example, in Figure \ref{F:N=7} it is shown the semiclassical graphs for the cycle with $N=7$ nodes. Since $N$ is prime, in this case the graph cannot be broken in more than single nodes, although we can also have different graphs. For $t_q = 1$ we recover the classical walk as expected, and the same for $t_q=6$. For $t_q = 2$ and $t_q=5$ node $0$ connects with nodes $2$ and $5$. But node $2$ connects to $4$, $4$ to $6$, $6$ to $1$, $1$ to $3$, $3$ to $5$, and $5$ to node $0$ again. So the graph is not broken. In this case it is again as a classical walk on the cycle with seven nodes, but the nodes are permuted. If we unroll the graph, it is similar to having a cycle formed by the chain of nodes $0-2-4-6-1-3-5$. For $t_q = 3$ and $t_q=4$ something similar happens, but with a different order of the nodes. Finally, for $t_q = N = 7$ the graph is broken into single nodes, closing a period in the quantum time. We can check that relation \eqref{n_graphs} holds true, having 4 different semiclassical walks in this case.

\begin{figure}[htpb]
	\centering
	\includegraphics[scale=0.5]{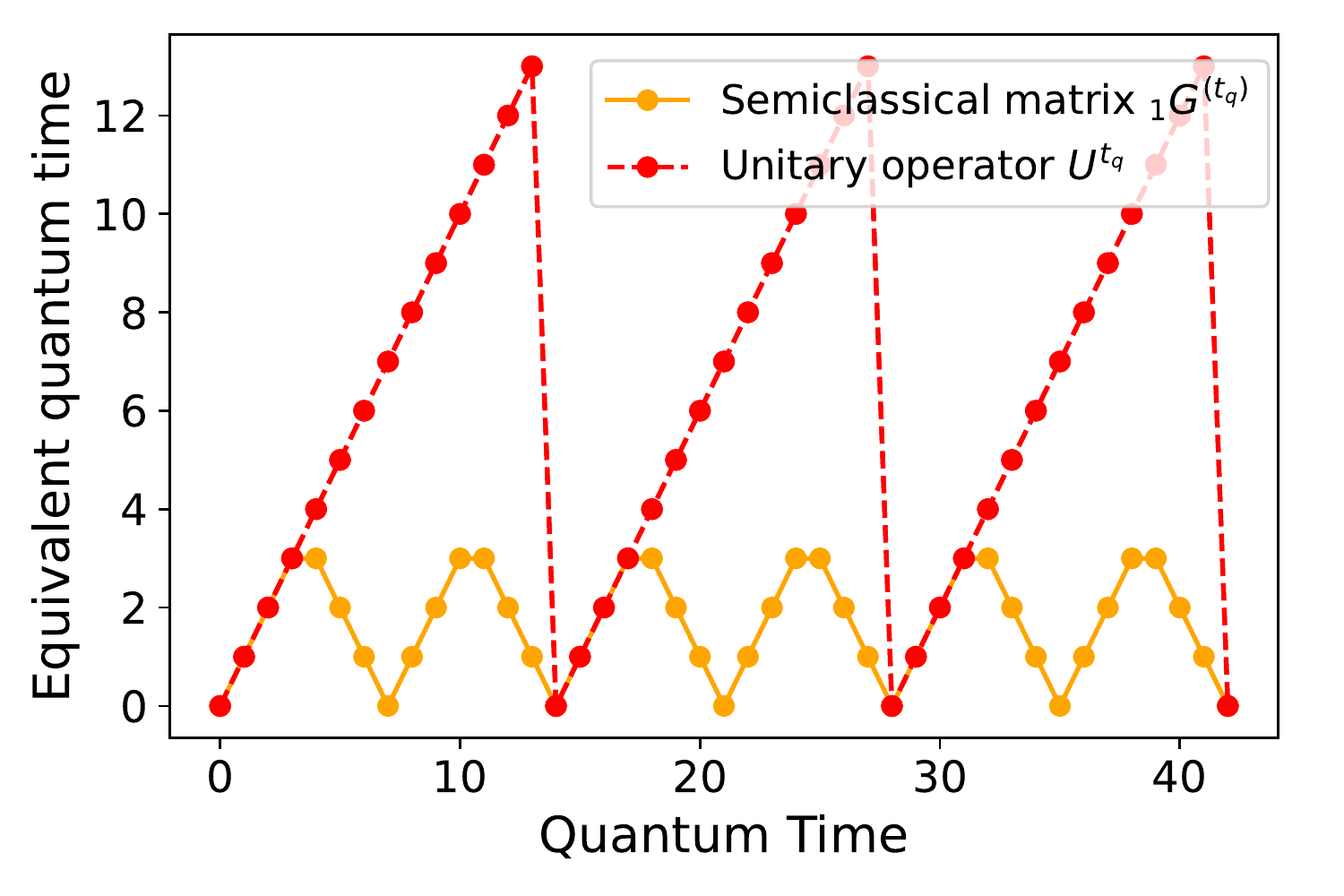}
	\caption{Periodicity of the semiclassical matrices and the unitary evolution for the cycle with $N=7$ nodes. At each quantum time it is represented the minimum value of $t_q$ for which there is a matrix that is equal. For example, for $t_q=10$ the semiclassical matrix is equal to the one at $t_q=3$. However, the unitary operator is not still repeated, so that the equivalent quantum time is also $t_q=10$. Time $t_q=0$ is not an actual walk, but is used to represent that the matrix is equal to the identity.}
	\label{F:N=7_period}
\end{figure}

In Figure \ref{F:N=7_period} it is shown the periodicity of the semiclassical matrices and the quantum evolution operator $U$ with the quantum time. It is clearly seen that the period of the semiclassical walks is $N=7$, having only 4 different semiclassical matrices. However, in this case the period of the unitary operator is $14$ instead of $7$. This is due to that despite $U$ having a period of $N$ when it acts over the set of states $\left|\psi_i\right>$, this set only generates a $N$ dimensional subspace of the entire $N^2$ dimensional Hilbert space. Since $U\left|\psi_i\right> = S\left|\psi_i\right>$, $U$ also has a period of $N$ over the set of the swapped states $S\left|\psi_i\right>$. Let us define:
\begin{equation}
	\mathcal{I}_U := \text{lin}\left\lbrace\left|\psi_i\right>,S\left|\psi_i\right>\right\rbrace,
\end{equation}
as the space generated by all the proxy states and their swapped states. Any vector $\left|a\right>$ in the orthogonal complement of $\mathcal{I}_U$ is perpendicular to both $\left|\psi_i\right>$ and $S\left|\psi_i\right>$. Thus, $\Pi\left|a\right> = 0$ and the first application of $U$ yields $U\left|a\right> = -S\left|a\right>$. Since $\left|a\right>$ is perpendicular to the states $S\left|\psi_i\right>$, then $S\left|a\right>$ is perpendicular to the states $\left|\psi_i\right>$ and $\Pi S\left|a\right>=0$. Thus, a second application of $U$ yields:
\begin{equation}
U^2\left|a\right> = -US\left|a\right> = S^2\left|a\right> = \left|a\right>,
\end{equation}
since $S^2 = \mathbbm{1}$, so the period of $U$ over $\mathcal{I}_U^\perp$ is just $2$. The total period of the unitary $U$ will be the least common multiple of the periods in both subspaces. Thus, for $N$ even the period is $N$, whereas for $N$ odd the period is $2N$.

We have shown only the results for $N=6$ and $N=7$ nodes. More results on different 1D cycles can be found in SM \cite{SM}.

\begin{figure}[htpb]
	\centering
	\subfigure[]{\includegraphics[scale=0.6]{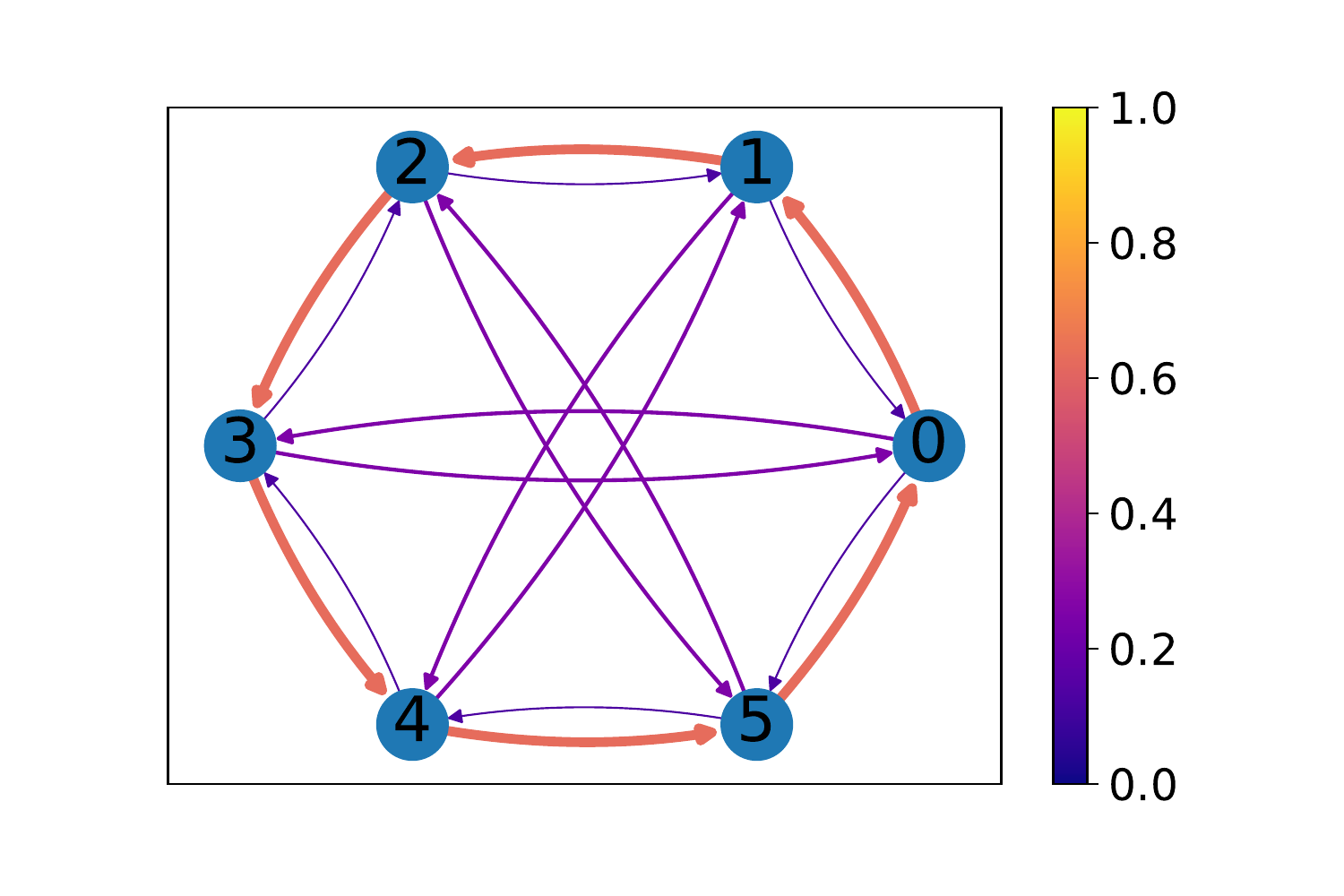}\label{F:isotropic}}
	\subfigure[]{\includegraphics[scale=0.6]{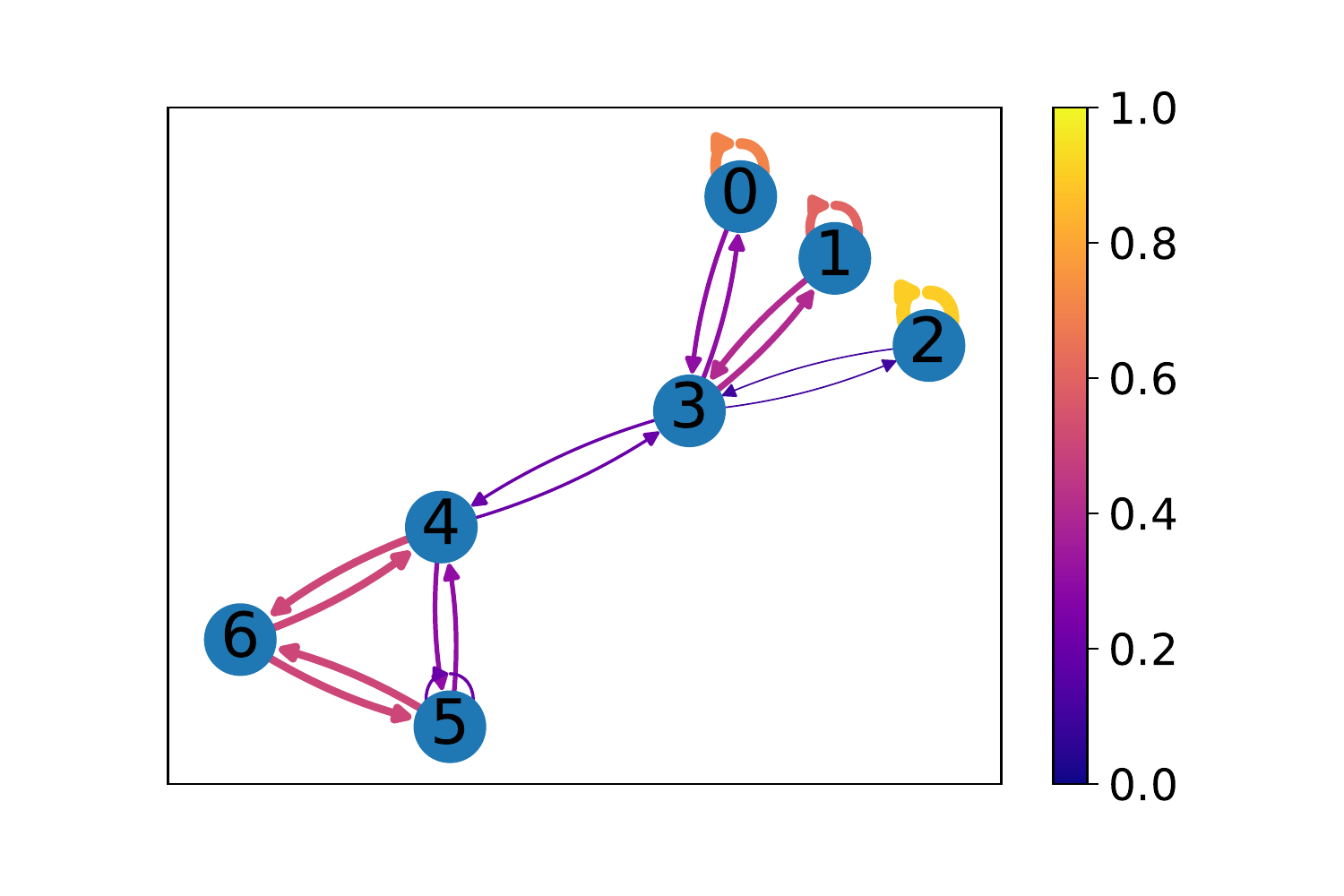}\label{F:symmetric}}
	\caption{a) Asymmetric and homogeneous graph. Between each pair of nodes the weights can be different in both directions, so it is asymmetric. However, all nodes have the same connectivity pattern, with weights that only depend on the relative distance between the nodes. b) Symmetric and inhomogeneous graph. Between each pair of nodes the weights are the same in both directions, so it is symmetric. However, each node has a different connectivity pattern, so it is inhomogeneous. The graphs have been plotted using the python library NetworkX \cite{NetworkX}.}
	\label{F:...}
\end{figure}

\section{Inhomogeneity-Driven Symmetry Breaking}\label{Symmetry}

Let us introduce the following two concepts:

-\textit{Symmetric graph}: it is a weighted graph whose transition matrix is symmetric, meaning that between each pair of nodes the probability for going from one to the other is the same in both directions. If $G$ is the transition matrix, then $G=G^T$.

-\textit{Homogeneous graph}: it is a weighted graph whose transition matrix elements only depend on the relative position of the nodes. Thus, all nodes have the same connectivity pattern and the same weights in their links.

In Figure \ref{F:isotropic} it is shown an instance of asymmetric and homogeneous graph. On one hand, the transition matrix is not symmetric since the probability for going from node $0$ to node $1$ is bigger than from node $1$ to node $0$. On the other hand, each node has the same behavior, meaning that the weights of their links depend only on the relative distance to the other nodes. In this case, each node $i$ connects with nodes $i \pm 1$ and $i + 3$, and the weights are the same for each node $i$. Thus, the graph is homogeneous.

In the examples of 1D cycles all the semiclassical graphs are symmetric. This could be due to the fact that the classical graphs were also symmetric. However, they were also homogeneous. So we wonder what happens when the classical graph is symmetric but inhomogeneous. With that purpose, we have built the graph shown in Figure \ref{F:symmetric}. It can be seen that the weights between each pair of nodes have the same intensity, so the transition matrix is symmetric. Nevertheless, each node has different connectivity patterns. For example, node $3$ connects to four nodes, whereas node $0$ only connects to one node and itself with a loop. Moreover, nodes $0$, $1$ and $2$ have also different weights in their links with node $3$ and the self-loop.

\begin{figure*}
	\centering
	\subfigure[]{\includegraphics[scale=0.375]{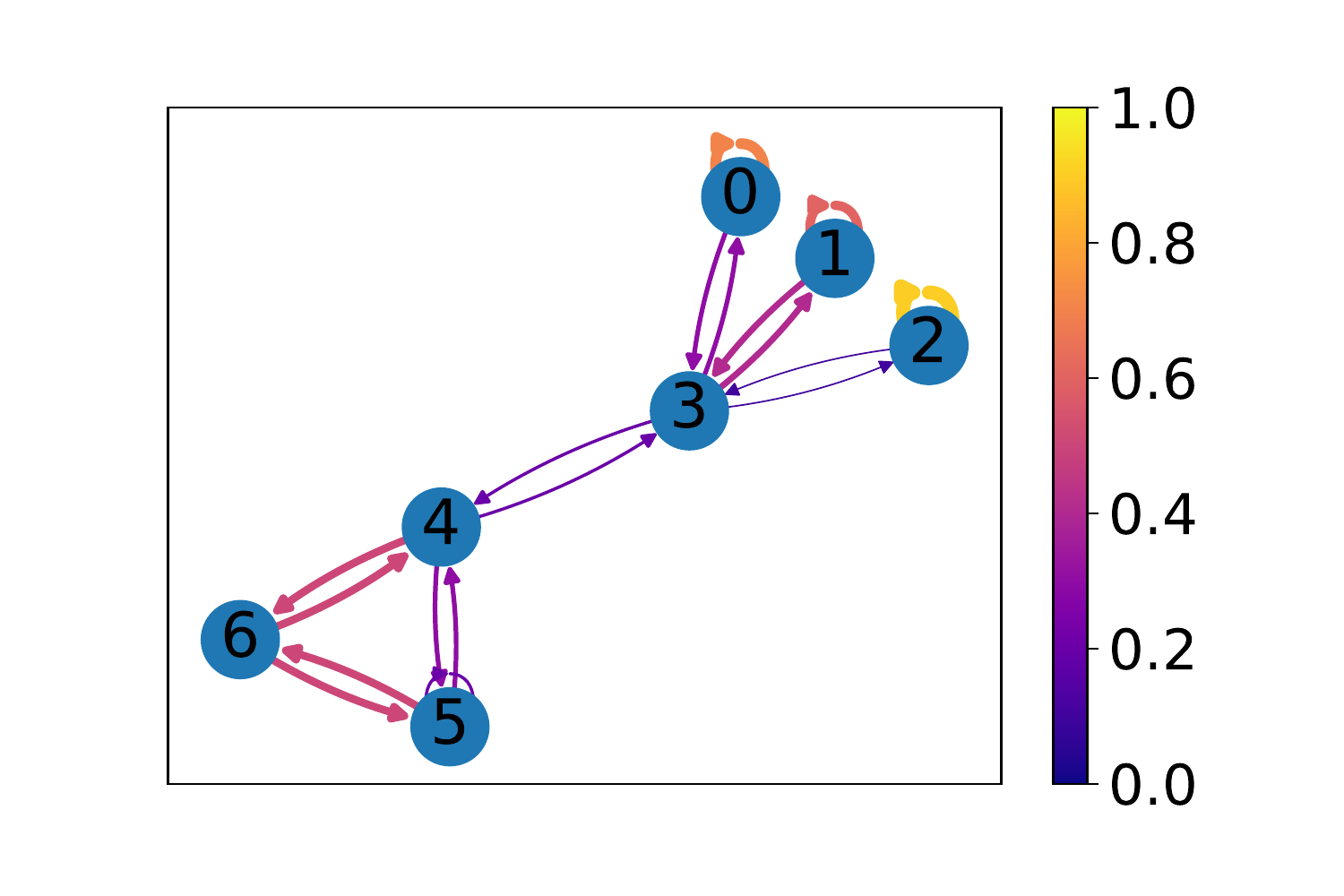}}
	\subfigure[]{\includegraphics[scale=0.375]{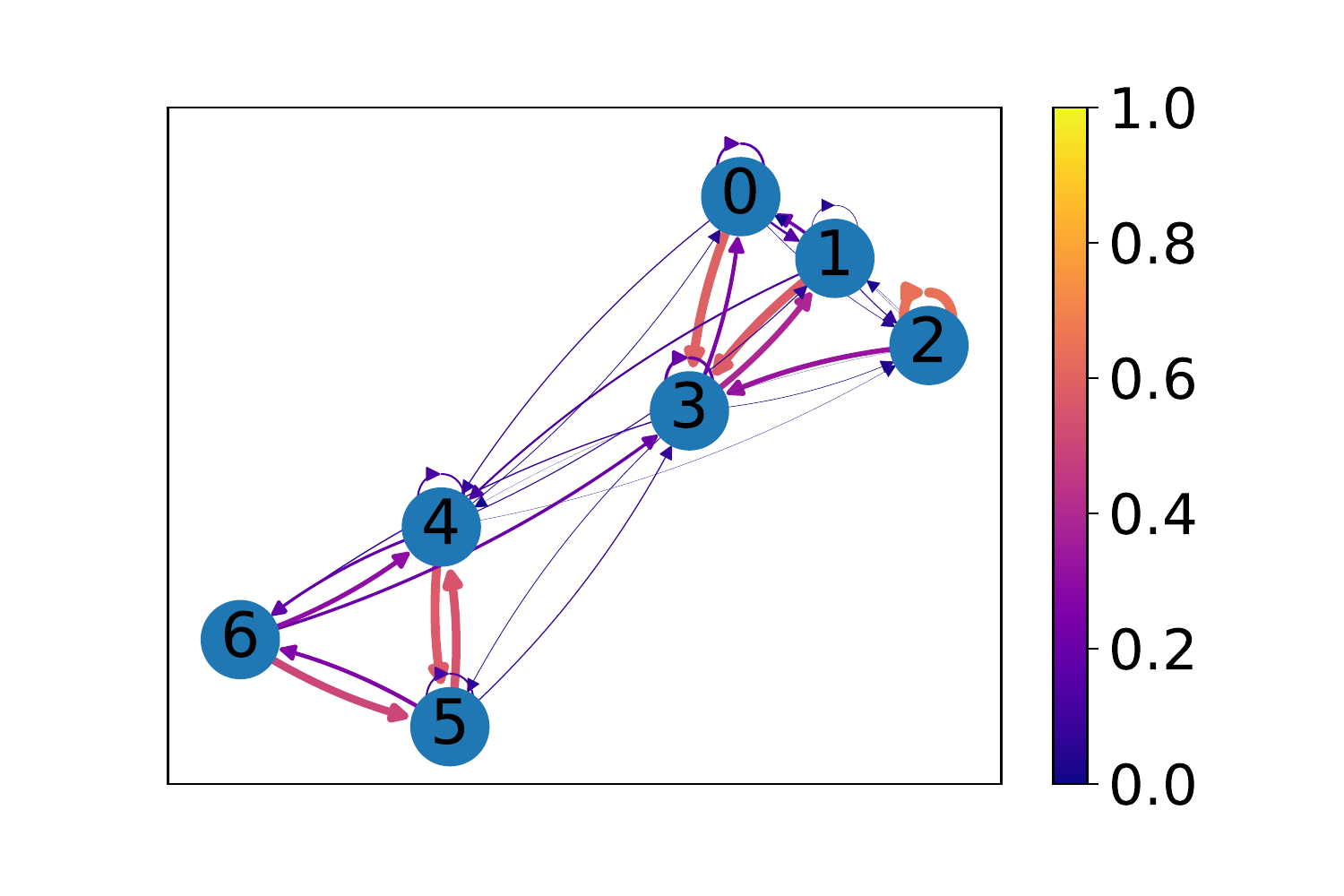}}
	\subfigure[]{\includegraphics[scale=0.375]{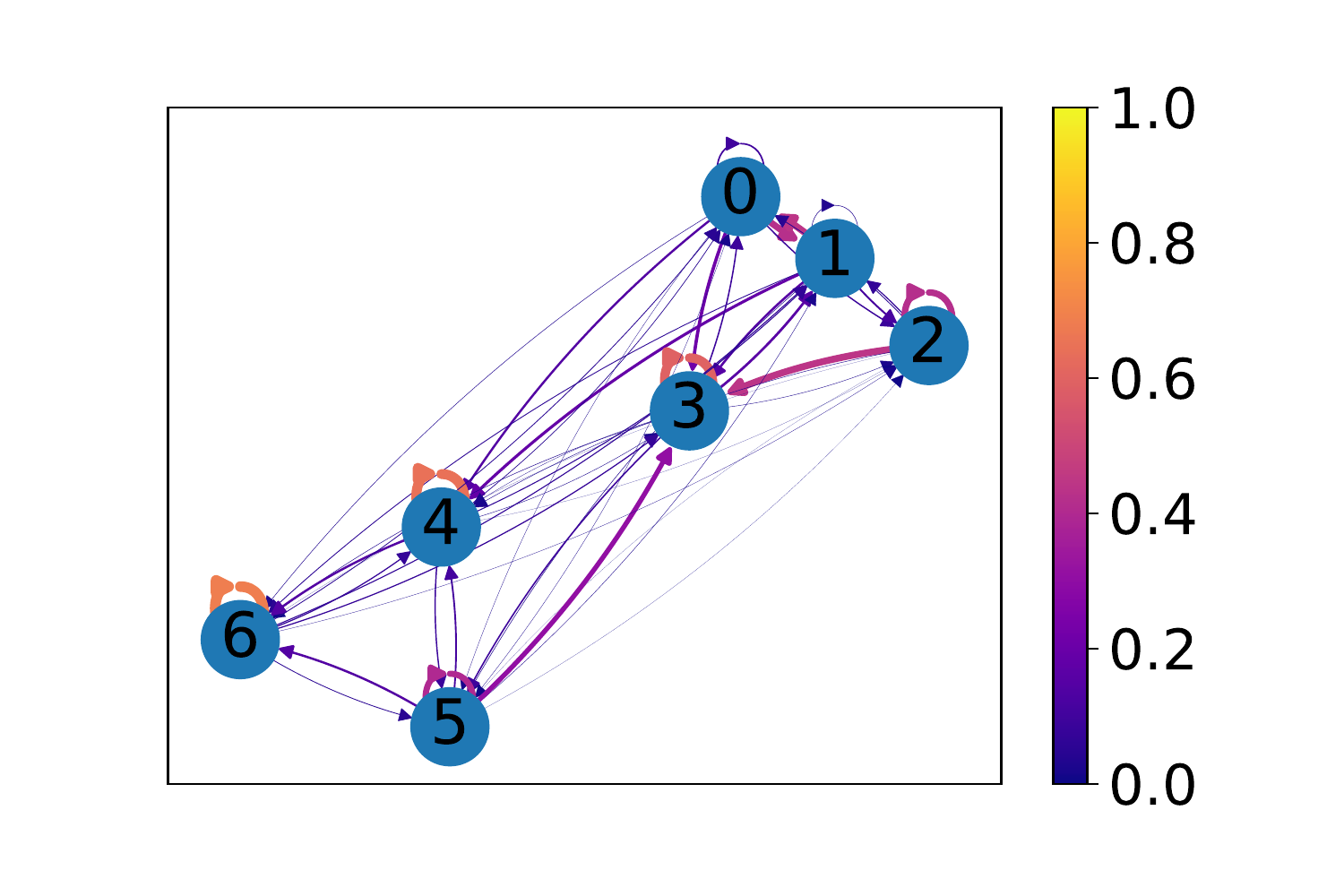}}
	\subfigure[]{\includegraphics[scale=0.375]{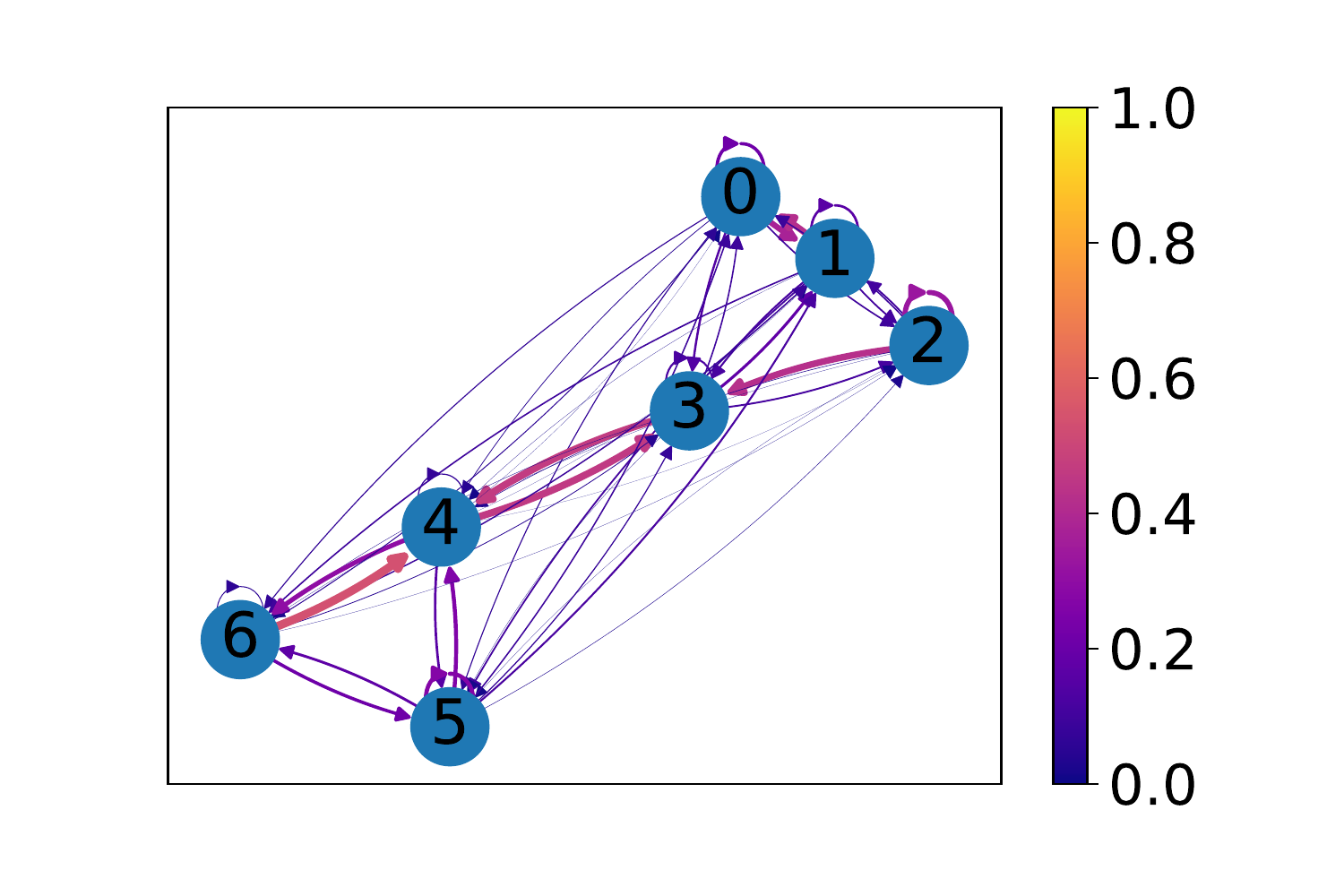}}
	\subfigure[]{\includegraphics[scale=0.375]{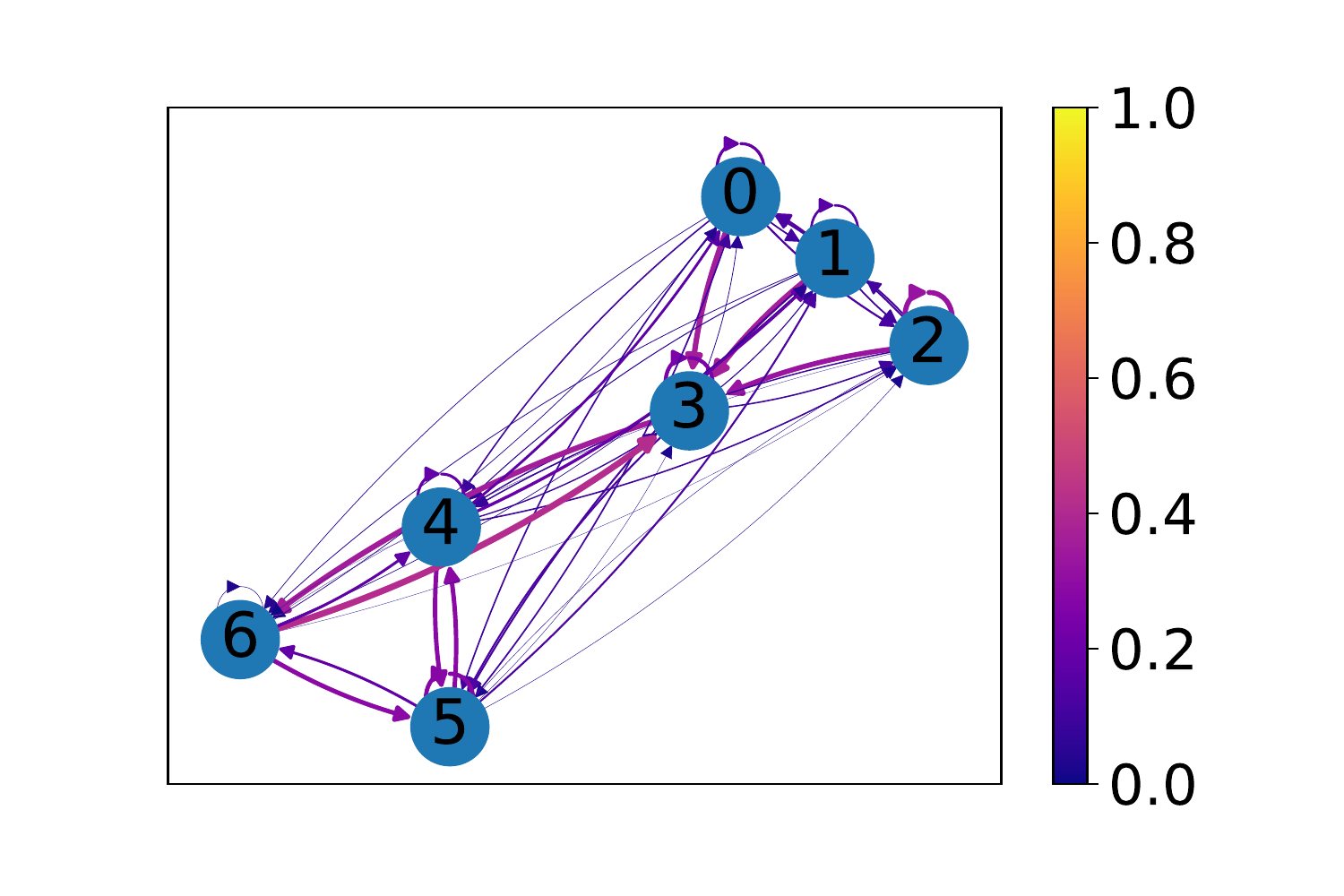}}
	\subfigure[]{\includegraphics[scale=0.375]{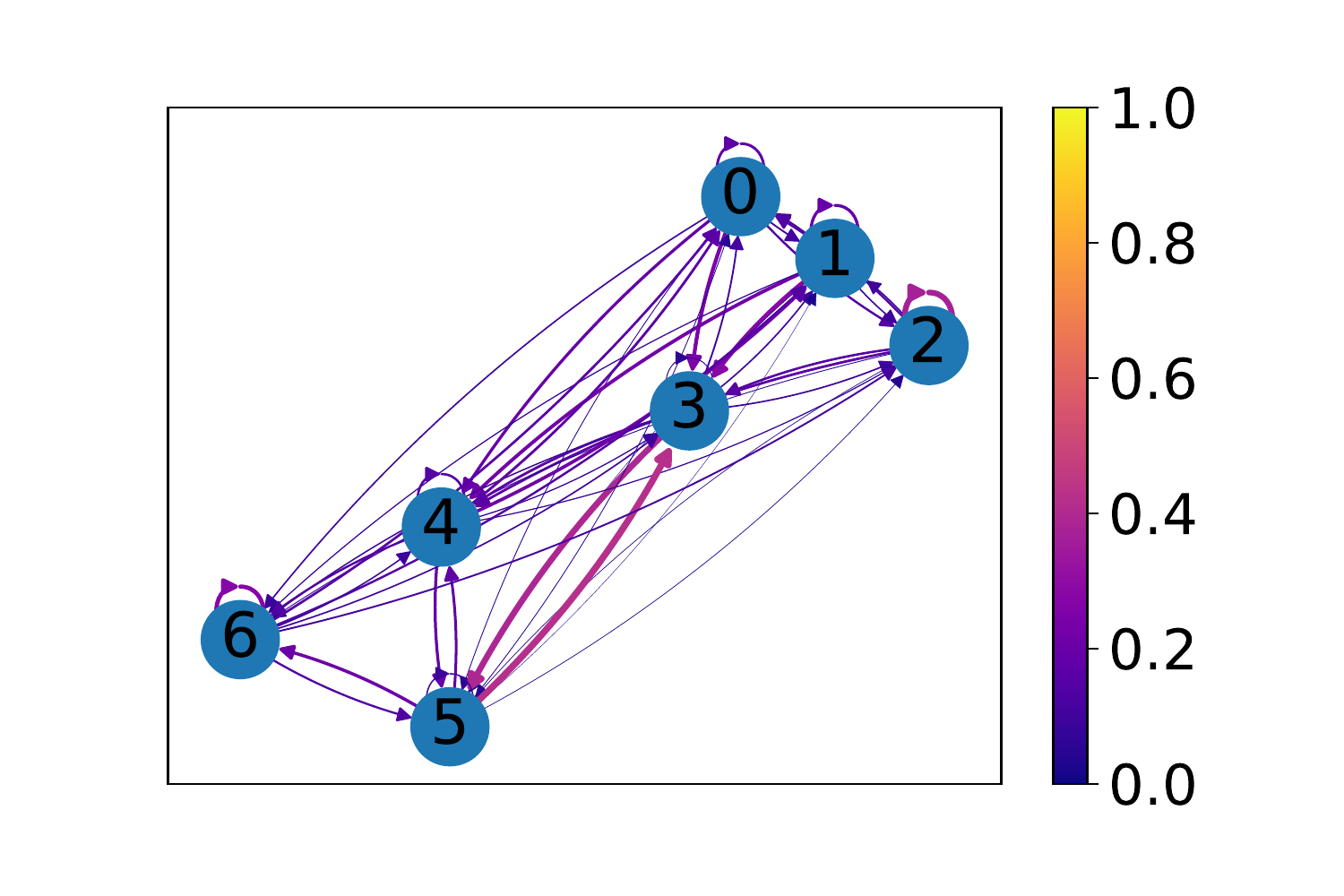}}
	\caption{Semiclassical graphs for the graph of Figure \ref{F:symmetric} for a) $t_q=1$, b) $t_q=2$, c) $t_q=3$, d) $t_q=4$, e) $t_q=5$, f) $t_q=6$. The weights of the edges are represented by the colormap. Moreover, to ease the visualization, the width of the edges are proportional to their weights. It can be seen how the symmetry of the graph is broken from $t_q=2$ onwards. The graphs have been plotted using the python library NetworkX \cite{NetworkX}.}
	\label{F:N=7_anisotropic}
\end{figure*}

We have simulated the semiclassical walks of the inhomogeneous graph in Figure \ref{F:symmetric}, and the first six semiclassical graphs are shown in Figure \ref{F:N=7_anisotropic}. For the first quantum time, $t_q=1$, we obtain the same as the classical graph, which is symmetric. However, for any other quantum time we observe that the symmetry has been broken. For example, for the graph with $t_q=2$ note that the weight in the edge that goes from node $6$ to $5$ is stronger than the weight from node $5$ to $6$. Moreover, since this is a more general case than with 1D cycles, there is not a periodicity in the semiclassical family. We have made simulations with other homogeneous symmetric graphs constructed at random finding that the symmetry is never broken, so we can conclude that the inhomogeneity is the cause of the symmetry breaking.

The fact that the semiclassical transition matrices are not symmetric anymore opens an application of the semiclassical walk to the problem of ranking nodes. Let us formulate the following theorems.

\textbf{Theorem 4: Uniform distribution for symmetric walks.} For a symmetric transition matrix $G=G^T$, the uniform distribution is an eigenvector with eigenvalue $1$, i.e., it is at equilibrium \cite{Th4}.

\textbf{Proof:} Let $G$ be the transition matrix, so that $G_{ij} = G_{ji}$, and let $\textbf{v}$ be the uniform probability vector, so that $\textbf{v}_i = 1/N \ \forall \ i$. We apply the transition matrix to this vector:
\begin{eqnarray}
[G\textbf{v}]_i &=& \sum_{j=0}^{N-1} G_{ij} \textbf{v}_j = \sum_{j=0}^{N-1} G_{ij} \frac{1}{N}\nonumber\\
&=& \frac{1}{N} \sum_{j=0}^{N-1} G_{ji} = \frac{1}{N} = \textbf{v}_i,
\end{eqnarray}
where we have used that $G$ is column-stochastic so that each column adds up to one \eqref{column_stochastic}. \qedsymbol

\textbf{Theorem 5: Uniform distribution for symmetric Quantum Szegedy's Walks}: Let G be a symmetric transition matrix, and $U$ the associated Szegedy unitary operator. Then, the uniform linear combination of all the $\left|\psi_i\right>$ states, denoted as
\begin{equation}\label{Psi_0}
\left|\Psi^{(0)}\right> = \frac{1}{\sqrt{N}}\sum_{i=0}^{N-1}\left|\psi_i\right>,
\end{equation}
is an eigenvector of $U$ with eigenvalue $1$.

\textbf{Proof:}
Since $\left|\Psi^{(0)}\right>$ is a linear combination of the $\left|\psi_i\right>$ states, the action of $U$ over it is just $S$. So, using the fact that $G_{ij} = G_{ji}$,
$$U\left|\Psi^{(0)}\right> = S\left[\frac{1}{\sqrt{N}}\sum_{i,k=0}^{N-1} \sqrt{G_{ki}} \left|i\right>_1\left|k\right>_2\right]$$
$$= \frac{1}{\sqrt{N}}\sum_{i,k=0}^{N-1} \sqrt{G_{ki}} \left|k\right>_1\left|i\right>_2 = \frac{1}{\sqrt{N}}\sum_{i,k=0}^{N-1} \sqrt{G_{ik}} \left|k\right>_1\left|i\right>_2$$
\begin{equation}
= \frac{1}{\sqrt{N}}\sum_{k=0}^{N-1}\left|\psi_k\right> = \left|\Psi^{(0)}\right>. \qed
\end{equation}

The limiting distribution of the classical walk is used in the PageRank algorithm to rank the nodes of the graph \cite{Brin1,Brin2,Brin3,Google_book}. However, for a symmetric transition matrix, in the case that the walk converges, it will converge to the uniform distribution, so no useful information can be obtained. In the case of the quantum PageRank, it is used the Szegedy's quantum walk starting from the uniform state $\left|\Psi^{(0)}\right>$ in \eqref{Psi_0}, and the distributions for each time step are averaged since the unitary character of the evolution makes it oscillate instead of converge \cite{Paparo1,Paparo2}. However, for a symmetric classical transition matrix the state $\left|\Psi^{(0)}\right>$ is an eigenvector, so the uniform distribution is obtained again. Thus, neither the classical nor quantum PageRank algorithms can rank successfully the nodes of a symmetric graph.

\begin{figure}[htpb]
	\centering
	\subfigure[]{\includegraphics[scale=0.6]{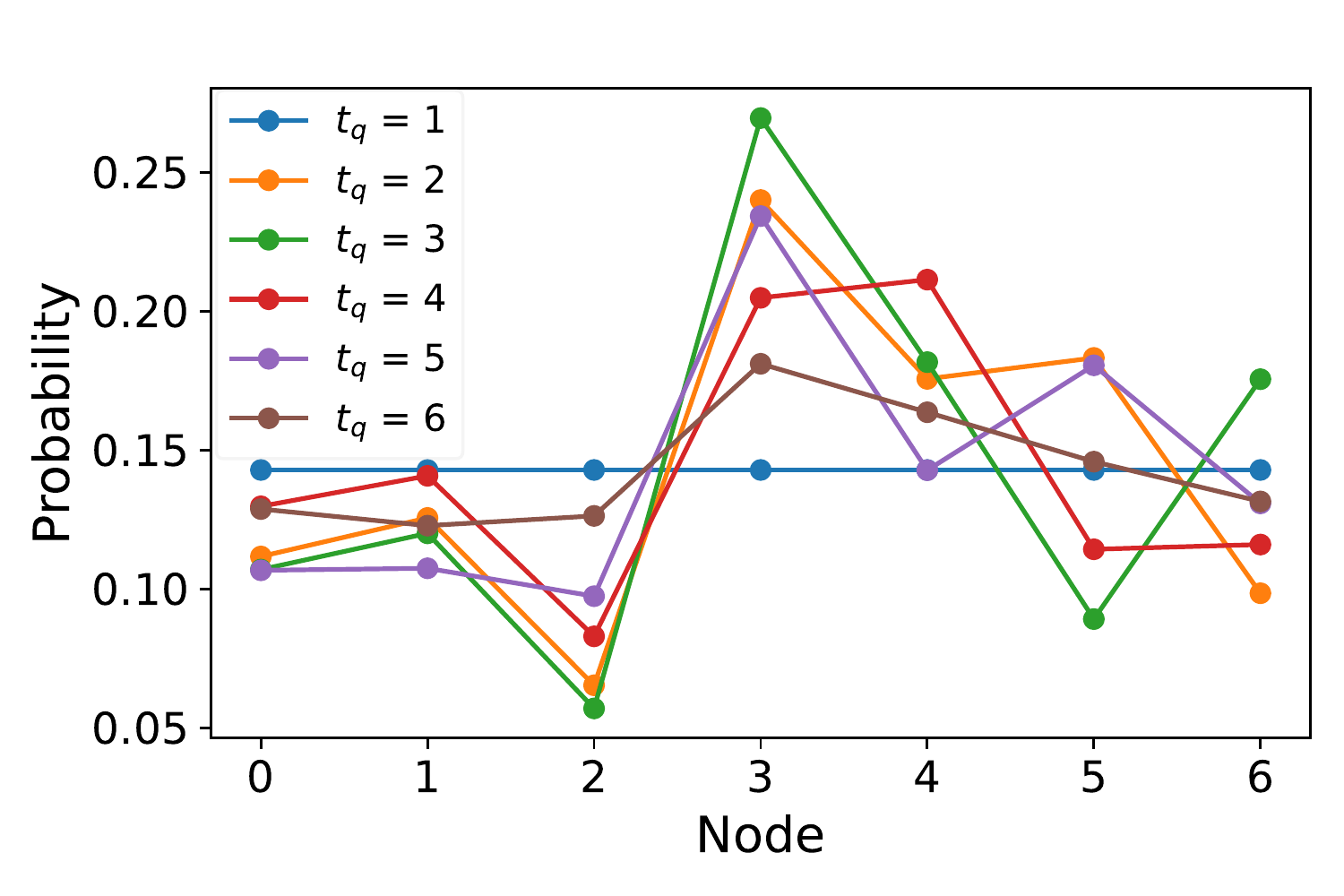}\label{F:PageRank_1}}
	\subfigure[]{\includegraphics[scale=0.6]{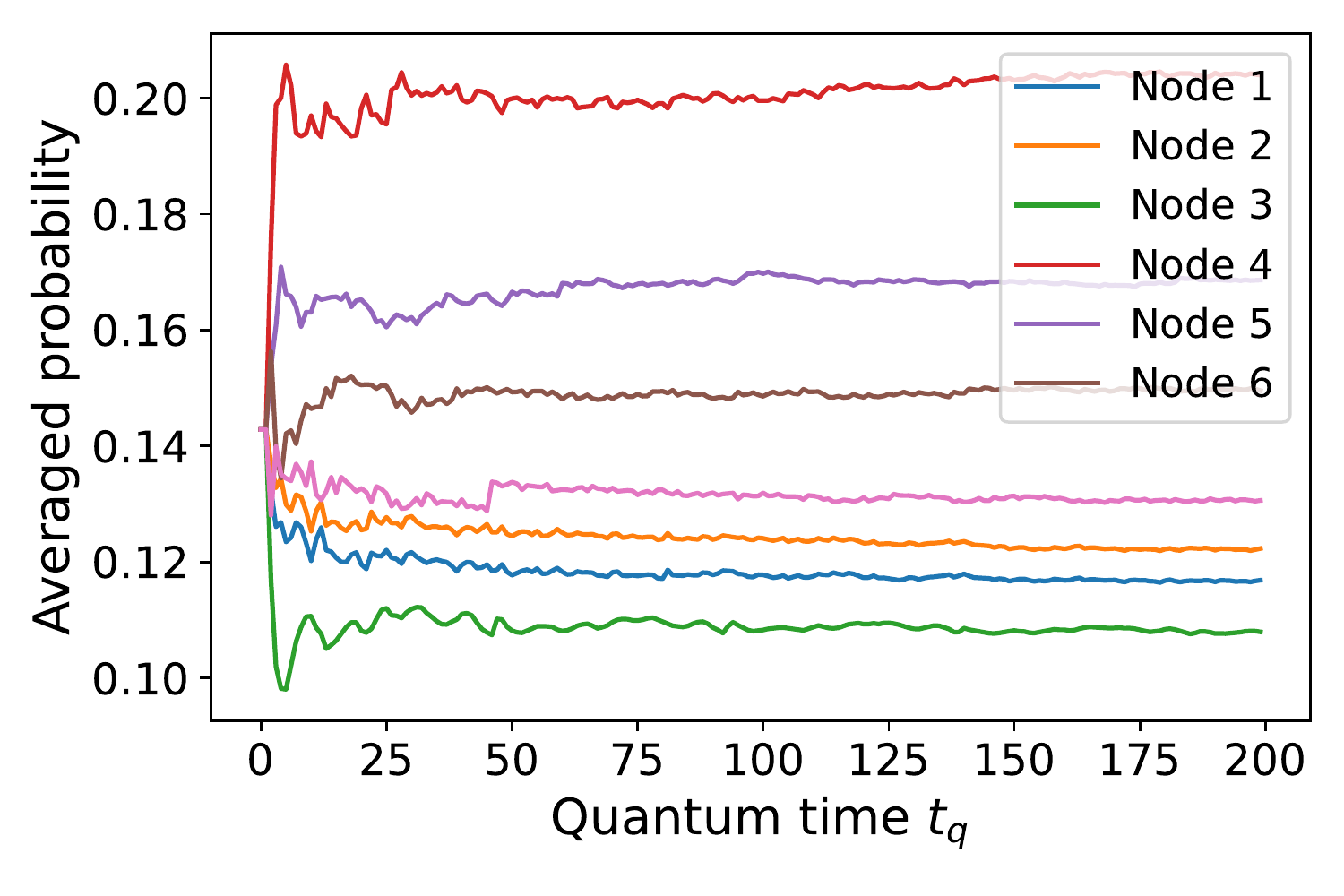}\label{F:quantum_convergence}}
	\subfigure[]{\includegraphics[scale=0.6]{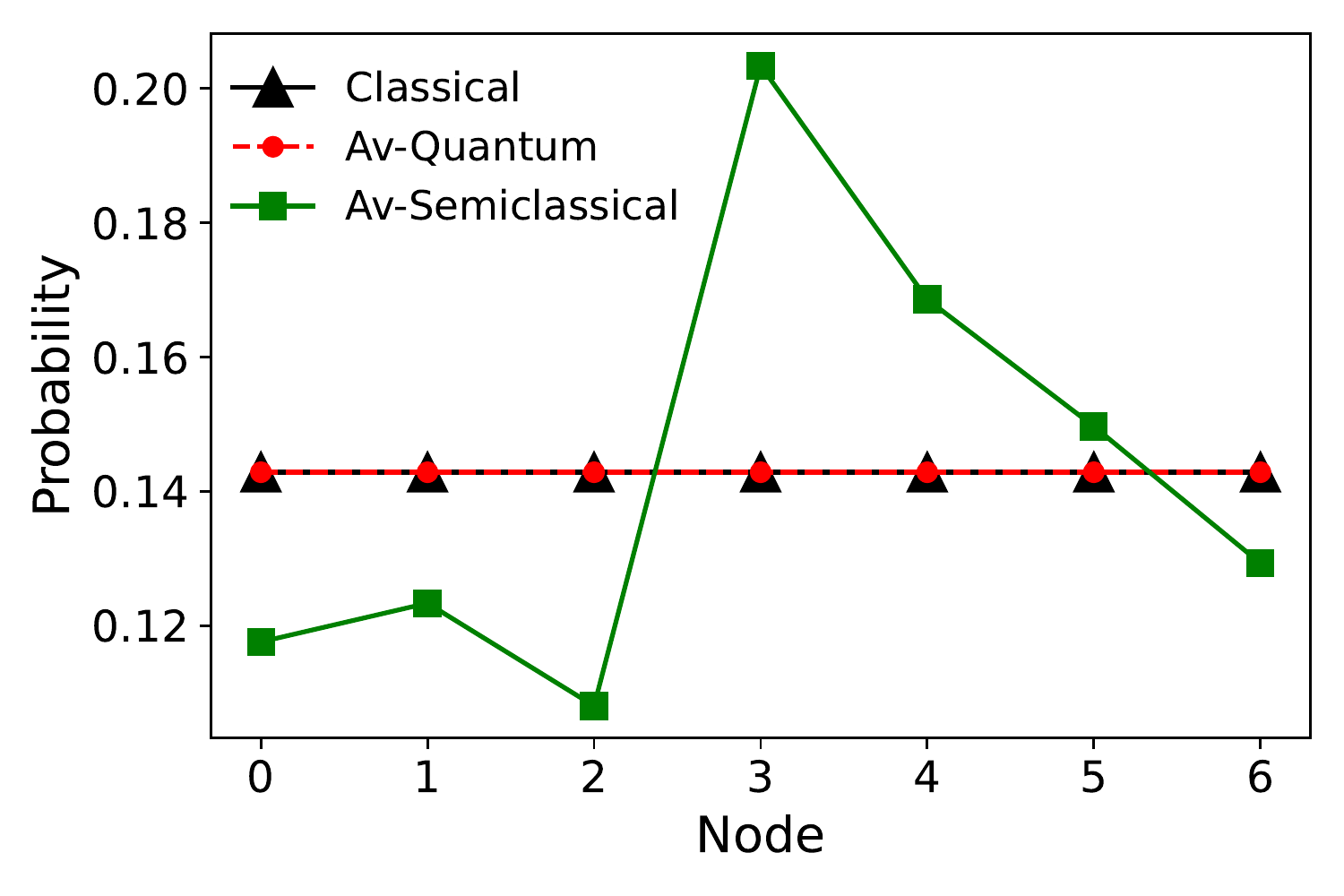}\label{F:PageRank_2}}
	\caption{Limiting distributions for the first six semiclassical walks of the graph in Figure \ref{F:symmetric}. b) Average of the limiting distributions vs. the quantum time used to average. c) Averaged limiting distribution of the semiclassical walks. It is compared with the classical and quantum PageRanks obtained from the classical graph, which yield a uniform distribution.}
	\label{...}
\end{figure}

Thanks to the symmetry breaking in the semiclassical graphs, the semiclassical walks converge to distributions different to the uniform one. The limiting distributions for the six semiclassical graphs of Figure \ref{F:N=7_anisotropic} are shown in Figure \ref{F:PageRank_1}. We now obtain different rankings for the nodes, but they are different for each semiclassical graph. To obtain an objective classification we average the distributions over the different quantum times, in a similar manner as the quantum PageRank does. For a Szegedy quantum walk it has been proved that averaging for a long enough time, the averaged distribution converges \cite{AvMCQW}. For the semiclassical walk, the same happens as can be seen in Figure \ref{F:quantum_convergence}. Finally, in Figure \ref{F:PageRank_2} it is shown the averaged distribution, compared with the uniform ones using the classical and quantum PageRank algorithms over the classical transition matrix $G$.

The final question is how this ranking relates to the structure of the network. Node $3$ is the most important being the node with more connections. The following most important is node $4$, which is the other central node. So it is as if the connectivity of the nodes plays a major role over the ranking. Furthermore, note that the differences in weights play also a role. Nodes $0$, $1$, and $2$, all linking to node $3$ and having a self-loop, are not degenerate due to the differences in the values of the weights. Node $1$ has the strongest weights in the edges with node $3$, and thus is the most important out of the three.

\section{Comparison with the continuous time approach}\label{MIQW}

The measurement-induced quantum walk \cite{MIQW} can be understood as a semiclassical walk where the quantum time is a continuous variable. The Hilbert space is the span of the computational basis $\left|i\right>$. Thus, in contrast with our approach, there is no need to reset any coin register after the measurement, since the system collapses to a suitable state for the next quantum evolution procedure. Thus, the transition matrices are obtained as:
\begin{equation}
	G^{(t_q)}_{ji} = \left|\left<j\right|U(t_q)\left|i\right>\right|^2 = \left|U_{ji}(t_q)\right|^2,
\end{equation}
where $U(t_q)$ is the unitary operator that performs the quantum evolution during a quantum time $t_q$. Note that despite the quantum time being continuous, the new transition matrix $G^{(t_q)}$ represents a walk in discrete classical time as in our approach.

In contrast to Szegedy's quantum walk, the unitary operator is obtained from the exponentiation of an hermitian operator $H$ related to the adjacency matrix of a graph, so that $U(t_q) = \exp(-iHt_q)$. Thus, whereas the Szegedy's quantum walk can be performed over any generic graph, in the continuous quantum walk there is the restriction that the graph must be undirected, so that the adjacency matrix is symmetric. Furthermore, since $H^T = H$, then the unitary matrix is also symmetric:
\begin{equation}
	U^T = e^{-iH^T} = e^{-iH} = U,
\end{equation}
and thus the transition matrices $G^{(t_q)}$ are also symmetric. This means that the uniform distribution is always a stationary state of the semiclassical walk. As the authors of \cite{MIQW} point out, in the case that the eigenvalue $1$ is not degenerate, it is as if the repeated measurements drove the system to a high temperature classical limit \cite{InfT}. This is not the case for our discrete-time semiclassical walks since we are not only measuring repeatedly, but also resetting to the desired proxy states using the classical information from the measurement. Thus, the reset scheme prevents this behavior.

\begin{figure*}
	\centering
	\subfigure[]{\includegraphics[scale=0.125]{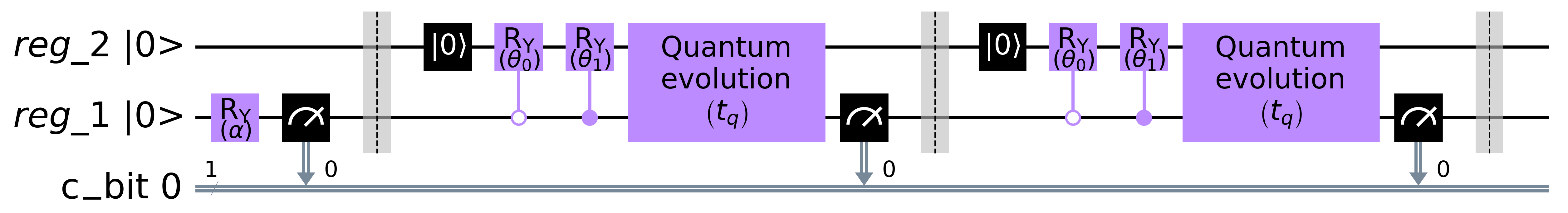}\label{F:circuit}}
	\subfigure[]{\includegraphics[scale=0.225]{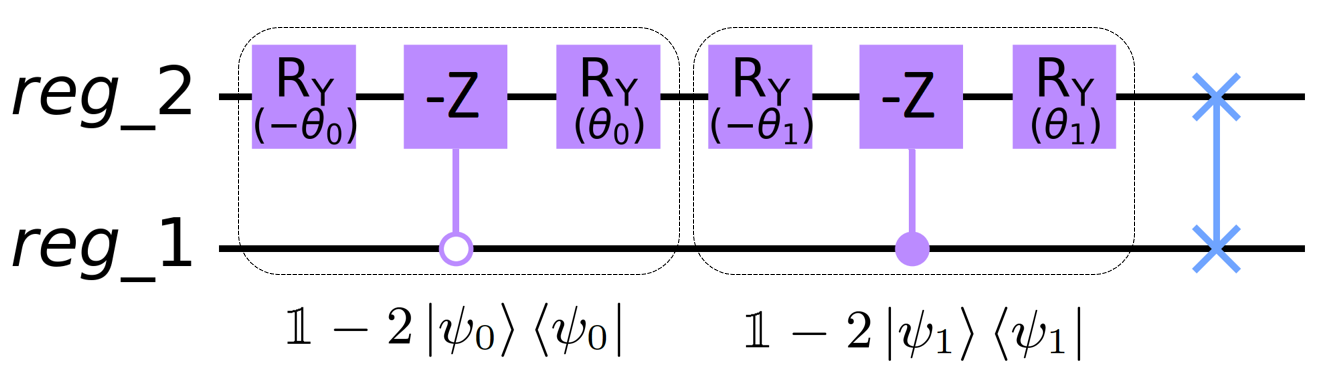}\label{F:quantum_evolution}}
	\caption{a) Quantum circuit for the semiclassical walks in the graph with two nodes whose classical transition matrix is given by \eqref{G_2_nodes}. This figure is an example with two classical steps, $t_c=2$, for a generic quantum time $t_q$. The values of the parameters are $\alpha = 0.927$, $\theta_0 = 2.5$ and $\theta_1 = 2.21$, b) Circuit for the unitary operator $U$. This operator can be expanded as $U = -S(\mathbbm{1} - 2\left|\psi_1\right>\left<\psi_1\right|)(\mathbbm{1} - 2\left|\psi_0\right>\left<\psi_0\right|)$. Moreover, using \eqref{rotation} we have that $\mathbbm{1} - 2\left|\psi_i\right>\left<\psi_i\right|=\left(\mathbbm{1} \otimes R_Y(\theta_i)\right)\left(\mathbbm{1}-2\left|i,0\right>\left<i,0\right|\right)\left(\mathbbm{1} \otimes R_Y(-\theta_i)\right)$. Finally, the gates $\mathbbm{1}-2\left|i,0\right>\left<i,0\right|$ can be implemented with controlled ($-Z$) gates, where $-Z = XZX$. The circuits have been plotted using the python library Qiskit \cite{Qiskit}.}
	\label{...}
\end{figure*}

Regarding the results in 1D cycles, the measurement-induced quantum walk is also able to break the graph into subgraphs. However, this breaking occurs for exceptional values of the quantum time, so they are very infrequent inside the family of semiclassical walks. Moreover, due to the continuous behavior of the quantum walk, except in the cases that the graph is broken, all the transition matrices are fully connected, meaning that there is a non-null probability for jumping from one node to any other or itself. Thus, the equivalent classical walk encoded by the hermitian matrix $H$ is never recovered in contrast to our discrete quantum time version.

\section{Experimental Semiclassical Walk on IBM-Q}\label{Experimental}

In order to show that the semiclassical walk functions in a real quantum processor unit (QPU), we have performed an experiment in the the platform IBM Quantum \cite{IBMQ}, with the processor ibmq-manila. This platform has been previously used to demonstrate experimentally other quantum walks such as the staggered quantum walk \cite{Staggered,Portugal_IBM}, and even the measurement-induced quantum walk \cite{MIQW_IBM}. Since current QPUs are very error prone, we have chosen a graph with only two nodes, where the Szegedy's quantum walk requires only two qubits, one per register of the state. The decoherence effects of quantum computers make the system end up in a uniform distribution. Thus, we have taken a weighted asymmetric graph so that the equilibrium distribution is different to the uniform one. The classical transition matrix of this graphs is
\begin{equation}\label{G_2_nodes}
G = \left(\begin{array}{cc}
	0.1 & 0.2\\
	0.9 & 0.8\\
\end{array}\right).
\end{equation}
In our experiment we wanted to obtain the probability distribution $p(t)$ of the walker being at each node for each classical time $t_c$. To do that, we must run the semiclassical walk several times to sample from the probability distributions, and then estimate the probabilities dividing the number of times the walker ends up in a node by the number of times the circuit has been run. Moreover, equation \eqref{classical_evolution} tells us that we can start from a non-trivial probability distribution $p(0)$. This can be achieved by initializing the circuit each time with a state taken at random from $p(0)$. To demonstrate that this actually works, we have taken the initial probability distribution as $p(0) = (0.8,0.2)^T$.

\begin{figure*}
	\subfigure[]{\includegraphics[scale=0.3875]{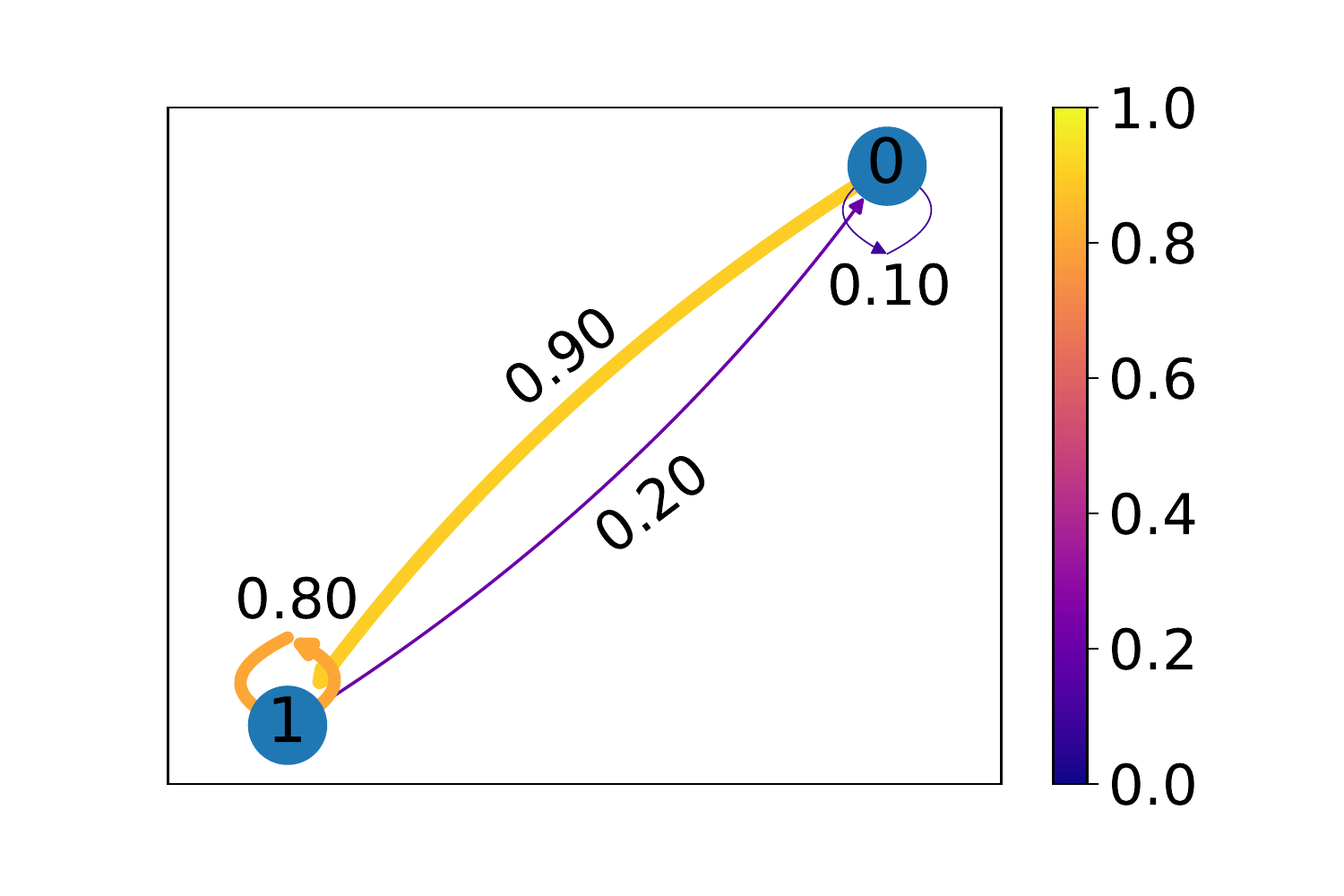}\label{F:G_N=2_tq=1}}
	\subfigure[]{\includegraphics[scale=0.3875]{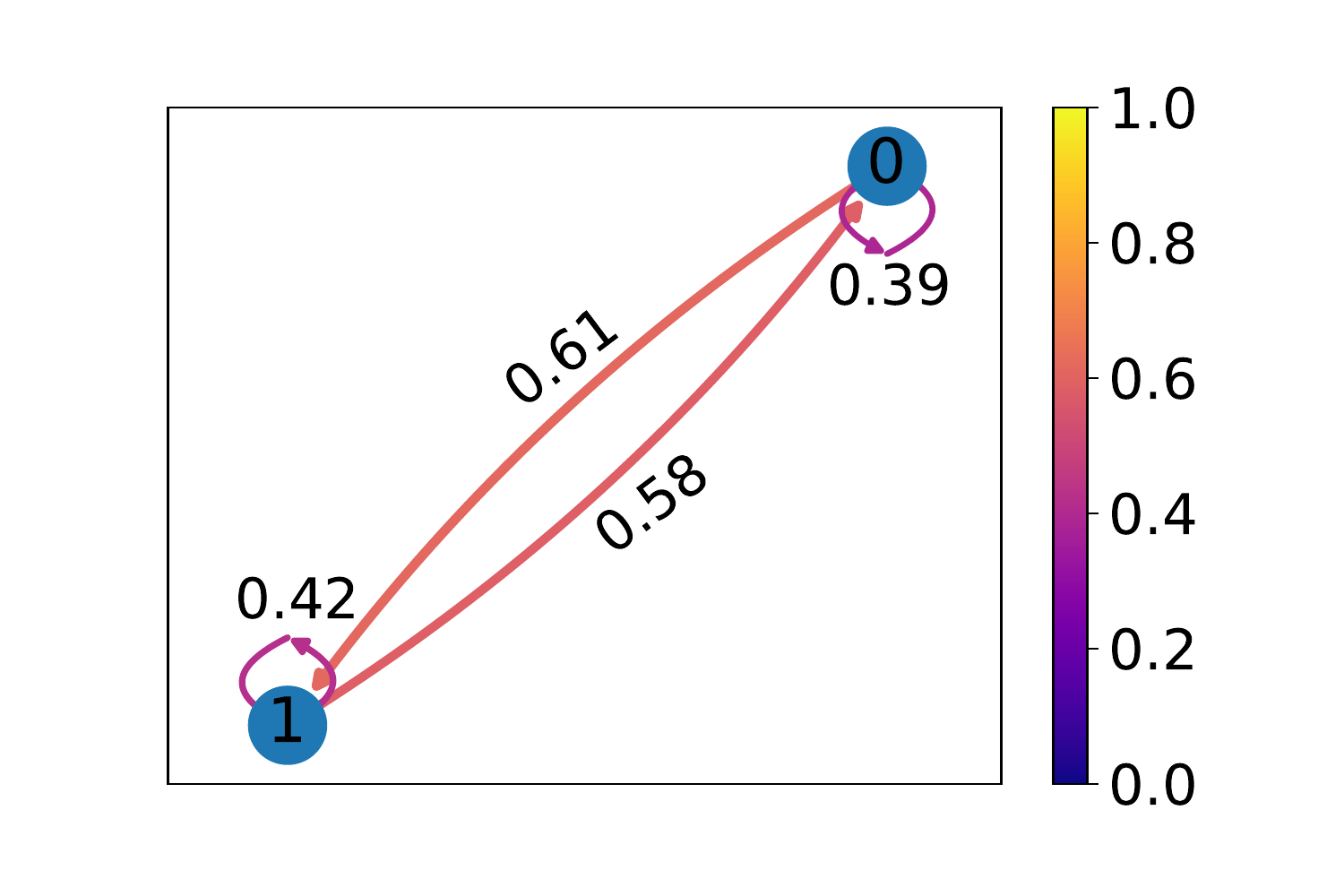}\label{F:G_N=2_tq=2}}
	\subfigure[]{\includegraphics[scale=0.3875]{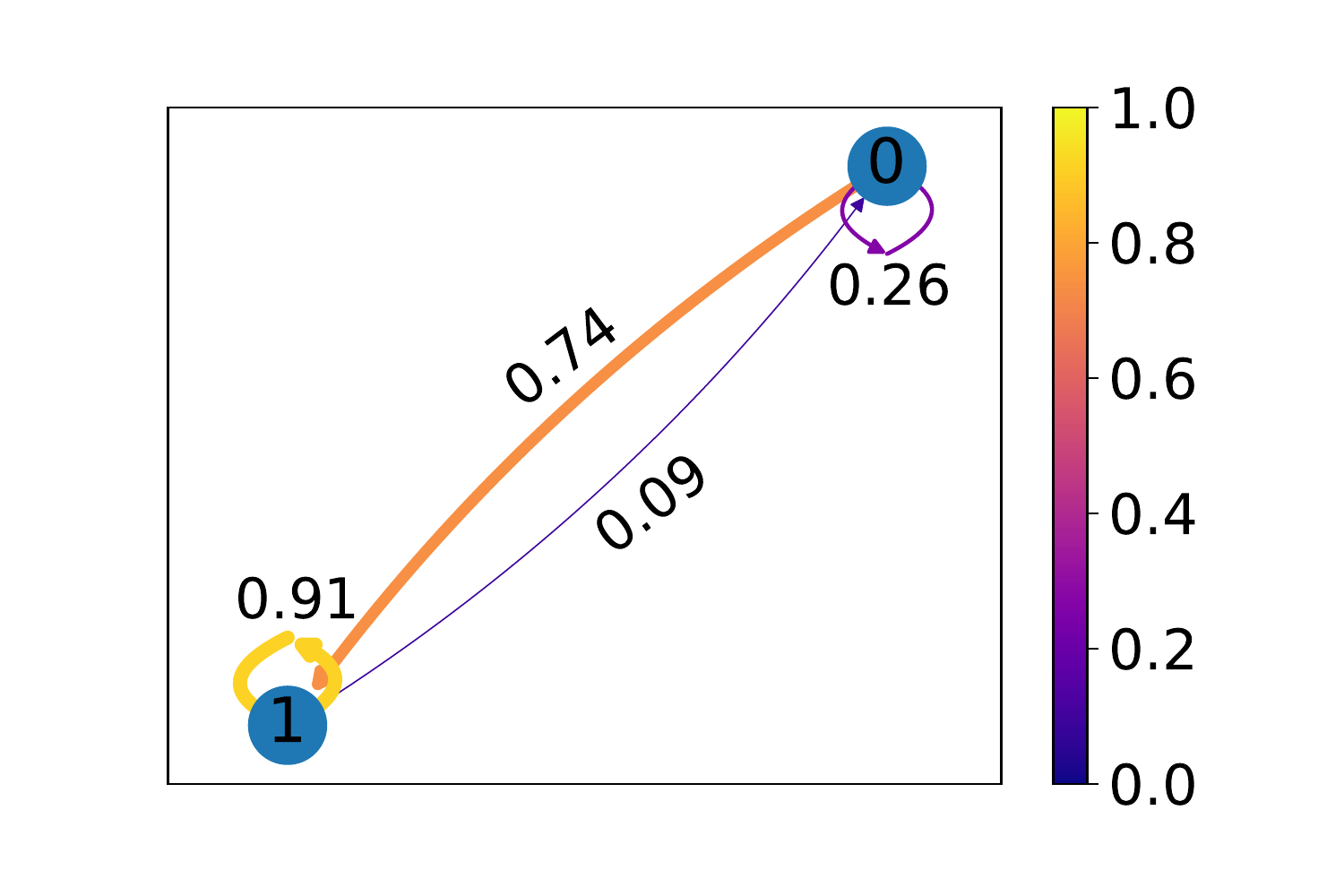}\label{F:G_N=2_tq=3}}
	\subfigure[]{\includegraphics[scale=0.375]{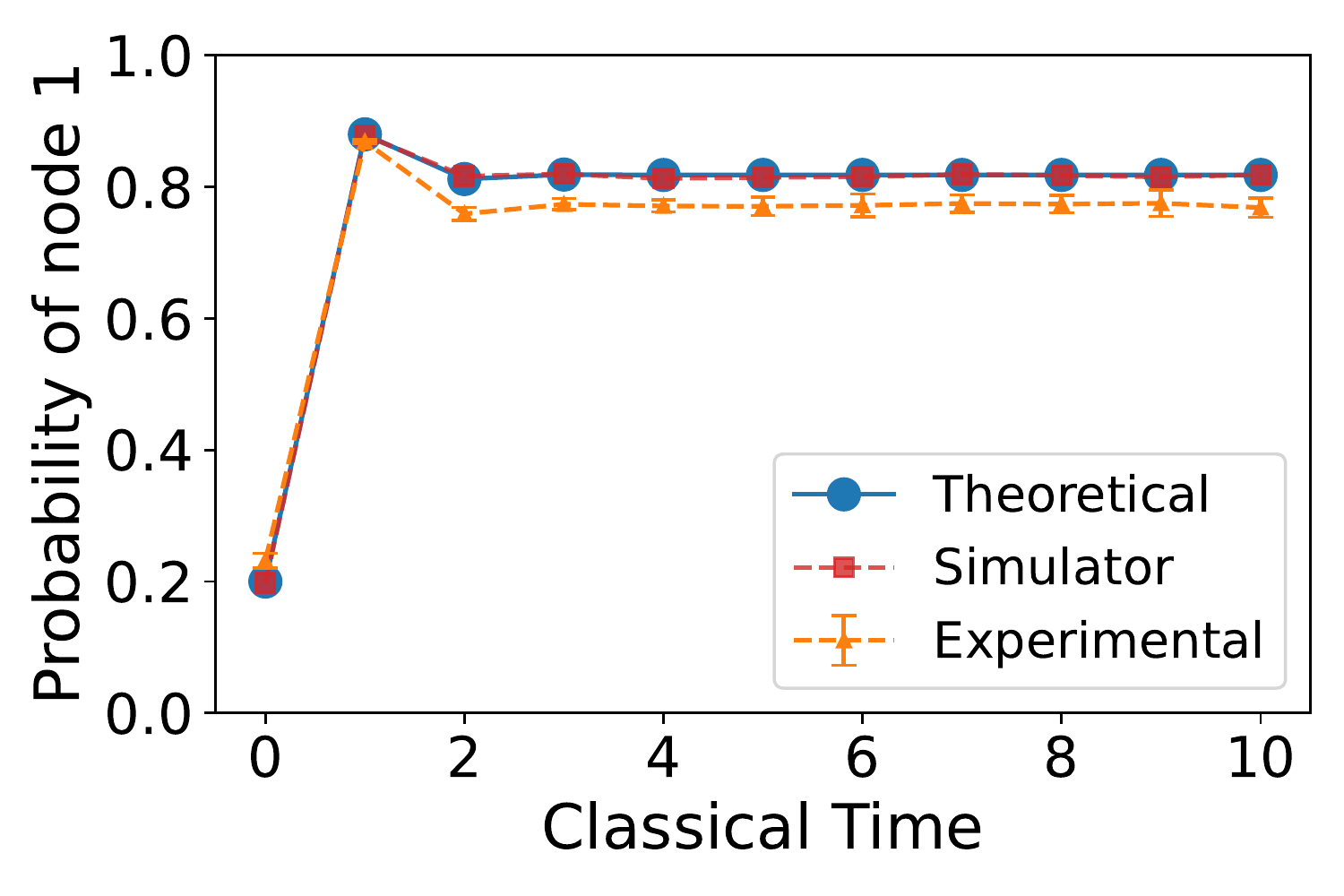}\label{F:exp_tq=1}}
	\subfigure[]{\includegraphics[scale=0.375]{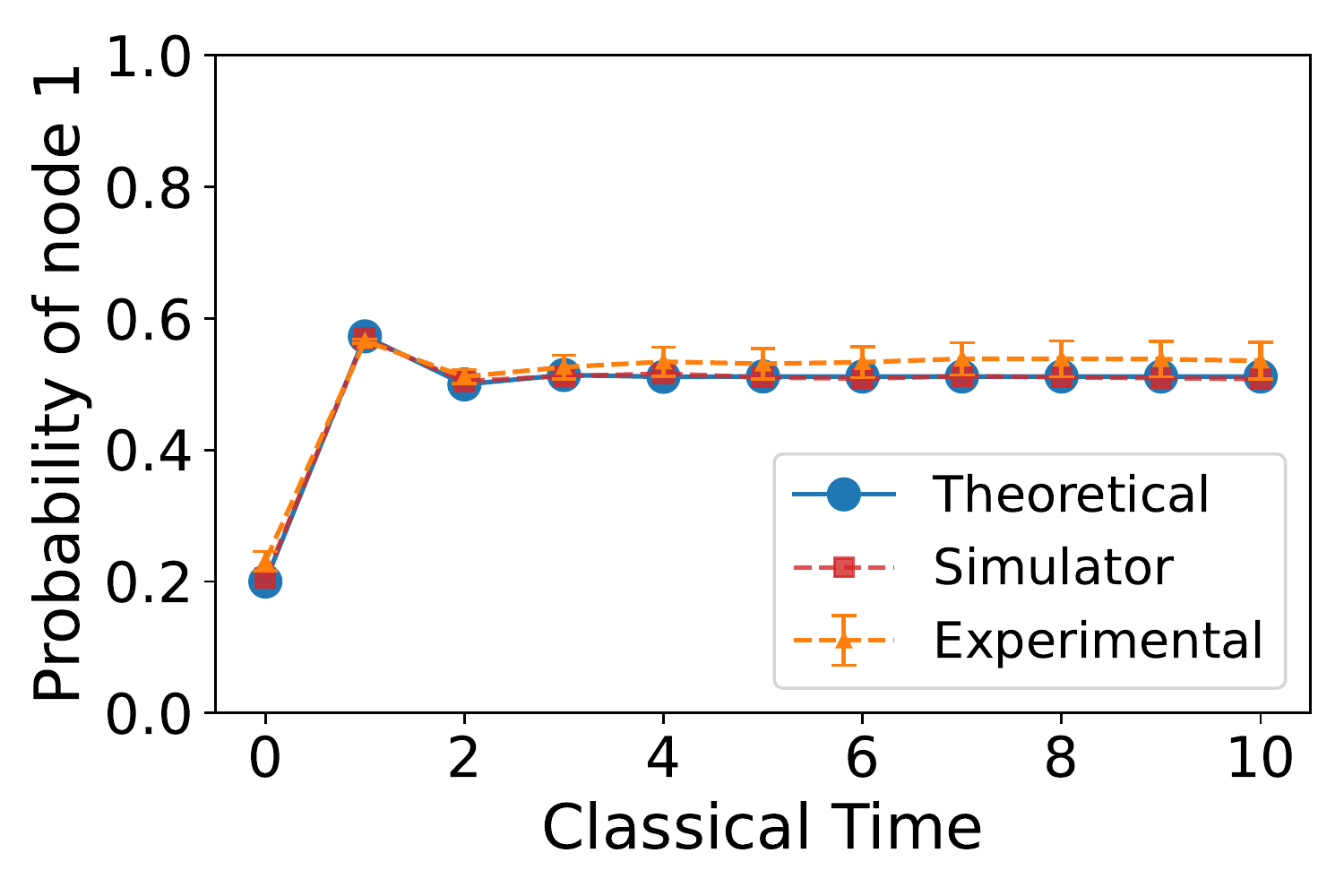}\label{F:exp_tq=2}}
	\subfigure[]{\includegraphics[scale=0.375]{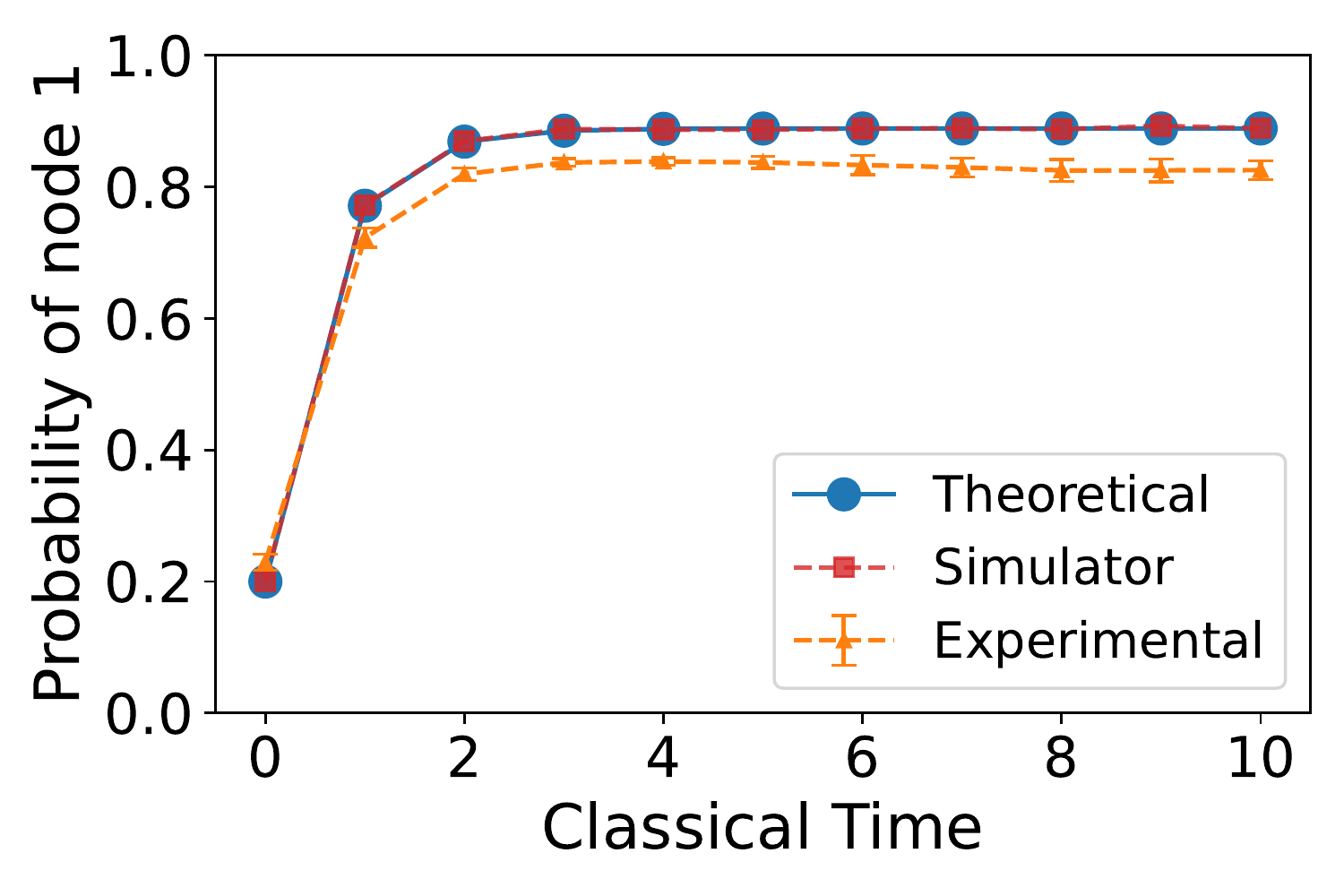}\label{F:exp_tq=3}}
	\caption{a)-c) Semiclassical graphs for the classical graph given by \eqref{G_2_nodes} for a) $t_q=1$, b) $t_q=2$, c) $t_q=3$. The weights of the edges are represented by the colormap and also over the edges. The graphs have been plotted using the python library NetworkX \cite{NetworkX}. d)-f) Experimental results for the semiclassical walks for d) $t_q=1$, e) $t_q=2$, f) $t_q=3$. Since there are only two nodes, the probability distribution can be represented by the probability of measuring node $1$. The error bars are computed with the standard deviations between ten different experiments. The results are compared with the theoretical results and the ones from the QASM simulator.}
	\label{...}
\end{figure*}

The general quantum circuit to perform the semiclassical walks is shown in Figure \ref{F:circuit}. The circuit always starts with both qubits in the state $\left|0\right>$. To randomly sample the initial distribution for taking an initial node, we apply a $R_Y(0.927)$ gate, so that the first register is converted to $\sqrt{0.8}\left|0\right> + \sqrt{0.2}\left|1\right>$. Thus, after measuring it, the first register is initialized to $\left|0\right>$ with an $80 \%$ of probability, and to $\left|1\right>$ with a $20 \%$ of probability. The result of this measurement corresponds to the position of the walker at $t_c=0$. For any other classical time, we repeat the following block. First, the second register is reset to $\left|0\right>$, which do not affect to the first register since they are not entangled due to the previous measurement. After that, we prepare the proxy state conditioned by the information of the first register, using controlled $R_Y(\theta_i)$ gates. We know that
\begin{equation}\label{rotation}
\left(\mathbbm{1} \otimes R_Y(\theta_i)\right) \left|i\right>_1\left|0\right>_2 = \left|\psi_i\right>,
\end{equation}
where $\theta_0 = 2.5$ and $\theta_1 = 2.21$. Thus, when the first register is in $\left|0\right>$ the proxy state $\left|\psi_0\right>$ is prepared, whereas when it is in $\left|1\right>$ the state $\left|\psi_1\right>$ is prepared instead. Once the proxy state is prepared, we apply the quantum evolution $U$ the number of quantum times $t_q$ required, where the circuit for the unitary $U$ is represented in Figure \ref{F:quantum_evolution}. Finally, we measure the state in the first register obtaining the position of the walker for that classical time. Note that we are using quantum controlled gates to reset the proxy states, which means that we would need one gate for each of the $N$ nodes in the graph. This would be very inefficient for complex graphs, so it would be far better to use the classical information of the measurement to tell the circuit what proxy state to prepare directly. Despite the fact that classically controlled gates are theoretically possible to implement, current QPUs of IBM do not allow them.

We have performed the experiments for the first three semiclassical walks of the family, $t_q=1,2,3$, whose semiclassical graphs are shown in Figures \ref{F:G_N=2_tq=1}-\ref{F:G_N=2_tq=3}. In each case, we have performed ten independent experiments and averaged the results. For each experiment the probability distributions were obtained sampling the circuit $20000$ times. The results are shown in Figures \ref{F:exp_tq=1}-\ref{F:exp_tq=3}. On one hand, in order to check the accuracy of the results, we can simulate the same circuit in the QASM simulator of Qiskit \cite{Qiskit}. This is a stochastic simulator that simulates a fault-tolerant quantum computer, also sampling a finite number of times to estimate the probability distributions. In our case we sample $20000$ times the circuits as in the real experiments. On the other hand, we can also calculate the theoretical probability distributions using the semiclassical matrices and equation \ref{classical_evolution}. In all cases the simulated circuits yield the same results as the theoretical formulation, verifying that the implemented circuit of Figure \ref{F:circuit} works.

The first semiclassical walk, for $t_q=1$, corresponds actually to the classical walk. We can see that the initial probability of the walker being at node $1$ is $0.2$. After the first step this probability rises to $0.88$. Then it goes down to $0.82$ and converges. The experimental results agree with the theoretical behavior. However, due to the errors of the real QPU it converges to approximately $0.77$, so the relative error is of a $6 \%$. For the second semiclassical walk, the graph is roughly symmetric, so it is not surprising that the limiting distribution is almost uniform. In this case the theoretical result converges to $0.51$ whereas the experimental one to $0.54$, so there is an approximate error again of a $6 \%$. Finally, for the third semiclassical walk the theoretical result converges to $0.89$ whereas the experimental one to $0.83$, with an error of a $7 \%$. Taking into account that current QPU are still very error prone, and we have not used any error mitigation nor error correction technique, our experimental results have an incredibly well agreement with the theoretical ones.

\section{Conclusions}\label{Conclusions}

We have reviewed the formulation of classical and quantum walks aiming at introducing the family of semiclassical walks. Each semiclassical walk can be seen as a classical walk where the transition matrix encodes a quantum evolution, and thus the dynamics is a mixture of classical and quantum evolution. Moreover, to be able to perform the semiclassical walk from quantum walks in discrete time, we have introduced the reset scheme and the proxy states, which encode the position of the walker in a graph.

We have formulated the semiclassical walk from the Szegedy's quantum walk, which is a walk in discrete time that is suitable for quantizing any classical transition matrix. The quantum states for Szegedy's quantum walk are formed by two registers, and each of both can yield information about the position of the walker. Thus, we have defined two classes of semiclassical Szegedy's walks depending on which register we use. Furthermore, we have proved that this two classes of semiclassical walks include the original classical walk, and there is an equivalence between them.

The semiclassical Szegedy's walks can be solved analytically for 1D lattices, such as cycles. We have obtained the semiclassical transition matrices for this case, and we have observed that for certain members of the semiclassical family the walk occurs in a broken graph. Moreover, there are some cases where the cycle is not broken, but the nodes are permuted. This results agree with the measurement-induced quantum walk, which is similar to the semiclassical walk but with a continuous quantum time, where the authors also observed that the 1D cycles could be broken.

Szegedy's quantum walk is suitable for any transition matrix, so it can be performed even in asymmetric graphs. Furthermore, we have observed that even if the classical graph is symmetric but not all nodes have the same behavior, i.e., it is not homogeneous, the symmetry of the transition matrix is broken in the semiclassical walk family. This can be useful for ranking nodes in graphs with symmetric transition matrices, where both classical and quantum PageRanks yield the uniform distribution. Due to the asymmetry in the semiclassical walks, these can converge to non-trivial distributions that can yield information about the graph. These results contrast with the ones of the continuous quantum time version, where all the semiclassical matrices are symmetric.

To demonstrate that the semiclassical walks can be implemented in quantum computers, we have performed some experiments in the platform IBM Quantum using a real QPU. We have used an asymmetric graph with $2$ nodes so that the limiting distributions are different to the trivial uniform one. We have done the experiments for the first three members of the semiclassical family, which include the classical walk. The results that we have obtained agree incredibly well with the theoretical ones, with a maximum error of a $7 \%$ due to the errors of current QPUs.

In the future, it would be interesting to apply these semiclassical Szegedy's walks to more complex graphs, in applications like optimization, quantum search or machine learning, where Szegedy's quantum walk has shown to have a good performance. Moreover, in contrast to quantum walks, the position of the walker can be measured at intermediate time steps. This could be crucial in algorithms that require knowing the position not only at the end of the walk.

\section*{Acknowledgments}

We acknowledge the support from the CAM/FEDER Project No.S2018/TCS-4342 (QUITEMAD-CM), Spanish MINECO grants MINECO/FEDER Projects,  PID2021-122547NB-I00 FIS2021, the “MADQuantum-CM" project funded by Comunidad de Madrid and by the Recovery, Transformation, and Resilience Plan – Funded by the European Union - NextGenerationEU and Ministry of Economic Affairs Quantum ENIA project. M. A. M.-D. has been partially supported by the U.S.Army Research Office through Grant No. W911NF-14-1-0103. S.A.O. acknowledges support from Universidad Complutense de Madrid - Banco Santander through Grant No. CT58/21-CT59/21.

\bibliography{MiBiblio}

\begin{thebibliography}{10}

\bibitem{QRW}
Y.~Aharonov, L.~Davidovich, and N.~Zagury.
\newblock {Quantum random walks}.
\newblock {\em Physical Review A}, 48:1687, 1993.

\bibitem{Trees}
E.~Farhi and S.~Gutmann.
\newblock {Quantum computation and decision trees}.
\newblock {\em Physical Review A}, 58:915, 1998.

\bibitem{Portugal}
R.~Portugal.
\newblock {\em {Quantum Walks and Search Algorithms}}.
\newblock New York: Springer, 2013.

\bibitem{Triangles}
F.~Magniez, M.~Santha, and M.~Szegedy.
\newblock {Quantum Algorithms for the Triangle Problem}.
\newblock {\em SIAM Journal on Computing}, 37:413--424, 2007.

\bibitem{ED}
A.~Ambainis.
\newblock {Quantum walk algorithm for element distinctness}.
\newblock {\em SIAM Journal on Computing}, 37:210--239, 2007.

\bibitem{QRW_Search}
N.~Shenvi, J.~Kempe, and K.~B. Whaley.
\newblock {Quantum random-walk search algorithm}.
\newblock {\em Physical Review A}, 67:052307, 2003.

\bibitem{Universal}
A.~M. Childs.
\newblock {Universal Computation by Quantum Walk}.
\newblock {\em Physical review letters}, 102:180501, 2009.

\bibitem{QSW}
J.~D. Whitfield, C.~A. Rodríguez-Rosario, and A.~Aspuru-Guzik.
\newblock {Quantum stochastic walks: A generalization of classical random walks
  and quantum walks}.
\newblock {\em Physical Review A}, 81:022323, 2010.

\bibitem{MIQW}
A.~Didi and E.~Barkai.
\newblock {Measurement-induced quantum walks}.
\newblock {\em Physical Review E}, 105:054108, 2022.

\bibitem{Szegedy}
M.~Szegedy.
\newblock {Quantum speed-up of Markov chain based algorithms}.
\newblock {\em 45th Annual IEEE Symposium on Foundations of Computer Science},
  pages 32--41, 2004.

\bibitem{Grover}
L.~K. Grover.
\newblock {A fast quantum mechanical algorithm for database search}.
\newblock {\em Proceedings of the 28th Annual ACM Symposium on Theory of
  Computing}, 1996.

\bibitem{Lemieux}
J.~Lemieux, B.~Heim, D.~Poulin, K.~Svore, and M.~Troyer.
\newblock {Efficient Quantum Walk Circuits for Metropolis-Hastings Algorithm}.
\newblock {\em Quantum}, 4:287, 2020.

\bibitem{Qfold}
P.~A.~M. Casares, R.~Campos, and M.~A. Martin-Delgado.
\newblock {QFold: Quantum Walks and Deep Learning to Solve Protein Folding}.
\newblock {\em Quantum Science and Technology}, 7:025013, 2022.

\bibitem{QMS}
R.~Campos, P.~A. Casares, and M.~A. Martin-Delgado.
\newblock {Quantum Metropolis Solver: QMS}.
\newblock {\em Quantum Machine Intelligence}, 5:28, 2023.

\bibitem{GWQMA}
G.~Escrig~Mas, R.~Campos, P.~A. Moreno~Casares, and M.~A. Martin-Delgado.
\newblock {Parameter Estimation of Gravitational Waves with a Quantum
  Metropolis Algorithm}.
\newblock {\em Classical and Quantum Gravity}, 40:045001, 2022.

\bibitem{Paparo1}
G.~D. Paparo and M.~A. Martin-Delgado.
\newblock {Google in a Quantum Network}.
\newblock {\em Scientific Reports}, 2:444, 2012.

\bibitem{Paparo2}
G.~D. Paparo, Müller M., F.~Comellas, and M.~A. Martin-Delgado.
\newblock {Quantum Google in a Complex Network}.
\newblock {\em Scientific Reports}, 3:2773, 2013.

\bibitem{APR}
S.~A. Ortega and M.~A. Martin-Delgado.
\newblock {Generalized quantum PageRank algorithm with arbitrary phase
  rotations}.
\newblock {\em Physical Review Research}, 5:013061, 2023.

\bibitem{Searchrank}
H.~Wang, J.~Wu, X.~Yang, P.~Chen, and X.~Yi.
\newblock {An Enhanced Quantum PageRank Algorithm Integrated with Quantum
  Search}.
\newblock {\em 2014 Eighth International Conference on Innovative Mobile and
  Internet Services in Ubiquitous Computing}, IEEE:74--81, 2014.

\bibitem{S_queries}
R.~A. Santos.
\newblock {Szegedy's quantum walk with queries}.
\newblock {\em Quantum Information Processing}, 15:4461--4475, 2016.

\bibitem{Paparo3}
G.~D. Paparo, V.~Dunjko, A.~Makmal, M.~A. Martin-Delgado, and H.~J. Briegel.
\newblock {Quantum Speedup for Active Learning Agents}.
\newblock {\em Physical Review X}, 4:031002, 2014.

\bibitem{Q_circuits}
T.~Loke and J.~B. Wang.
\newblock {Efficient quantum circuits for Szegedy quantum walks}.
\newblock {\em Annals of Physics}, 382:64--84, 2017.

\bibitem{Metropolis}
N.~Metropolis, A.~W. Rosenbluth, M.~N. Rosenbluth, A.~H. Teller, and E.~Teller.
\newblock {Equation of State Calculations by Fast Computing Machines}.
\newblock {\em The journal of chemical physics}, 21:1087--1092, 1953.

\bibitem{Hastings}
W.~K. Hastings.
\newblock {Monte Carlo sampling methods using Markov chains and their
  applications.}
\newblock {\em Biometrika}, 57:97--109, 1970.

\bibitem{SimAnn}
S.~Kirkpatrick, C.~D. Gelatt~Jr, and M.~P. Vecchi.
\newblock {Optimization by Simulated Annealing}.
\newblock {\em science}, 220:671--680, 1983.

\bibitem{Brin1}
S.~Brin and L.~Page.
\newblock {The anatomy of a large-scale hypertextual web search engine}.
\newblock {\em Computer Networks and ISDN Systems}, 30:107--117, 1998.

\bibitem{Brin2}
S.~Brin, R.~Motwami, L.~Page, and Winograd. T.
\newblock {What can you do with a web in your pocket?}
\newblock {\em IEEE Data Engineering Bulletin}, 21:37--47, 1998.

\bibitem{Brin3}
L.~Page, S.~Brin, R.~Motwami, and T.~Winograd.
\newblock {The PageRank citation ranking: Bringing order to the web}.
\newblock {\em Stanford InfoLab}, 1998.

\bibitem{Google_book}
A.~N. Langville and C.~D. Meyer.
\newblock {\em Google’s Pagerank and Beyond: The Science of Search Engine
  Rankings}.
\newblock Princeton university press, 2011.

\bibitem{QT}
{By quantum time we refer to a parameter that determines the duration of the
  quantum evolution, rather than a quantum operator.}

\bibitem{Notes}
A.~Childs.
\newblock {Quantum algorithms: LECTURE 14. Discrete-time quantum walk}.
\newblock {\em University of Waterloo}, 2008.

\bibitem{SM}
{See Supplementary Material for proofs of theorems 2 and 3, and for results in
  other 1D cycles. It can be found online at
  \url{https://doi.org/10.1016/j.physa.2023.129021}.}

\bibitem{NetworkX}
A.~Hagberg, D.~S~Chult, and P.~Swart.
\newblock Exploring network structure, dynamics, and function using networkx.
\newblock In G.~Varoquaux, T.~Vaught, and J.~Millman, editors, {\em Proceedings
  of the 7th Python in Science Conference}, pages 11 -- 15, Pasadena, CA USA,
  2008.

\bibitem{Th4}
A.~W. Marshall, I.~Olkin, and B.~C. Arnold.
\newblock {\em {Inequalities: theory of majorization and its applications.}}
\newblock Springer New York Dordrecht Heidelberg London, 1979.

\bibitem{AvMCQW}
R.~Balu, C.~Liu, and S.~E. Venegas-Andraca.
\newblock {Probability distributions for Markov chains based quantum walks}.
\newblock {\em Journal of Physics A: Mathematical and Theoretical}, 51:035301,
  2017.

\bibitem{InfT}
J.~Yi, P.~Talkner, and G.~L. Ingold.
\newblock {Approaching infinite temperature upon repeated measurements of a
  quantum system}.
\newblock {\em Physical Review A}, 84:032121, 2011.

\bibitem{Qiskit}
M.~S. Anis~et al.
\newblock {Qiskit: An Open-source Framework for Quantum Computing}, 2021.

\bibitem{IBMQ}
IBM Quantum.
\newblock {https://quantum-computing.ibm.com/}.
\newblock 2021.

\bibitem{Staggered}
R.~Portugal, R.~A. Santos, T.~D. Fernandes, and D.~N. Gonçalves.
\newblock {The Staggered Quantum Walk Model}.
\newblock {\em Quantum Information Processing}, 15:85--101, 2016.

\bibitem{Portugal_IBM}
F.~Acasiete, F.~P. Agostini, J.~K. Moqadam, and R.~Portugal.
\newblock {Implementation of quantum walks on IBM quantum computers.}
\newblock {\em Quantum Information Processing}, 19:1--20, 2020.

\bibitem{MIQW_IBM}
S.~Tornow and K.~Ziegler.
\newblock {Measurement induced quantum walks on an IBM Quantum Computer}.
\newblock {\em arXiv:2210.09941}, 2022.

\end{thebibliography}
\bibliographystyle{unsrt}

\end{document}